\documentclass[preprint,tightenlines,aps,showpacs,showkeys,amsmath,amssymb]{revtex4}
\usepackage{graphicx}
\begin{document} 

\title{Density reorganization in hot nuclei}

\author{S. K. Samaddar$^{1}$, J. N. De$^{1}$, X. Vi\~nas$^{2}$, and
M. Centelles$^{2}$}
\affiliation{
$^1$Saha Institute of Nuclear Physics, 1/AF Bidhannagar, Kolkata
{\sl 700064}, India \\
$^2$Departament d'Estructura i Constituents de la
Mat\`eria, Facultat de F\'{\i}sica, \\
Universitat de Barcelona, 
Diagonal {\sl 647}, {\sl 08028} Barcelona, Spain}


\begin{abstract}
The density profile of a hot nuclear system produced in intermediate
energy heavy ion collisions is studied in a microcanonical formulation
with a momentum and density dependent finite range interaction.
The caloric curve and the density evolution with excitation are
calculated for a number of systems for the equilibrium mononuclear
configuration; they compare favorably with the recent experimental
data. The studied density fluctuations are seen to build up rapidly
beyond an excitation energy of $\sim$ 8 MeV/u indicating the 
instability of the system towards nuclear disassembly. Explicit
introduction of deformation in the expansion path of the heated
nucleus, however, shows that the system might fragment even earlier.
We also explore the effects of the nuclear 
equation of state and of the mass and
isospin asymmetry on the nuclear equilibrium configuration and the 
relevant experimental observables.

\end{abstract} 



\maketitle

\section{Introduction}
 In intermediate energy nuclear collisions, a hot nuclear system is
initially produced in unstable equilibrium, generally in a compressed 
state. The unbalanced thermal and compressional pressure push the
system towards expansion, till an equilibrium configuration is reached.
If the system is sufficiently excited (above a few MeV per nucleon),
the mononuclear configuration is ultimately driven towards a subnuclear
density (break-up density $\rho_f$) below which it no longer remains 
in the said configuration and undergoes disassembly into a number of
fragments. 

  Study of the properties of hot nuclear systems at subnormal
densities is of utmost physical interest in understanding 
explosive nucleosynthesis in supernova explosions \cite{ish,bot}, it plays 
a decisive factor in the element composition in the heavenly bodies.
Nuclear multifragmentation studies in the laboratory serve as a 
window to have a closer look at these processes; the properties 
of the diluted nuclear material prior to fragmentation have 
a key role in such an understanding. It further helps in mapping
the nuclear equation of state (EoS) and also extracting, from
the complex experimental data, the density dependence of the 
nuclear symmetry energy \cite{li,she} that 
are crucial inputs in detailing the 
supernova explosion and discerning the properties of neutron
star matter.

 From experimental and theoretical studies, it is seen that
the break-up density $\rho_f$ decreases with increasing
excitation energy. Since this density has a profile, use of
a single value of it leaves some room for ambiguity. In
experimental studies, it is understood as an average density.
There are several experimental estimates of the break-up density. 
This has been determined from the studies of correlation
functions of emitted light particles from the source \cite{fri},
but the results seem to depend on the choice of the pair of
correlated light fragments. It raises the question on the
representation of the evolutionary stage of the reaction
process as given by the correlation data and thus there remains 
a  latitude for considerable uncertainties. The systematics of the
Coulomb barriers required to fit the intermediate mass ejectile
spectra also give an estimate of the break-up densities \cite{vio}.
However, in this extraction process, it is implicitly assumed 
that the barrier corresponds to a binary system \cite{rad}, 
contrary to the fact that the hot system disassembles into
many fragments. This introduces doubts in the proper estimation
of the break-up densities. Recently, the break-up densities
have also been derived from the analysis of apparent level
density parameters required to fit the measured caloric
curves \cite{nat}; the extraction, however, hinges on
a suitable mix of theory \cite{nor} with experiment. 
The deduced break-up densities from different sets of
experimental data have sometimes wide variations
among them, a sound knowledge
of these densities is thus still incomplete.

The disassembly of a mononuclear configuration at excitations
above $\sim 3 - 4$ MeV/u has been theoretically studied in
a statistical framework for more than the last two decades.
There have been a variety of these models \cite{bon,gro},
the basic underlying premise is that  
the excited nuclear system expands
to a {\it freeze-out} volume and then the system undergoes
a one-step {\it prompt multifragmentation}, the fragmentation
pattern being determined by the available phase space. It is
assumed that both thermal and chemical equilibrium are established
among the fragments at the freeze-out density $\rho_f$ so that
the fragmentation pattern is frozen out there. In most of these models,
the freeze-out density is independent of the excitation energy.
Different models, however, use widely varying freeze-out densities.
Whereas in the canonical or microcanonical models \cite{bon,gro},
the density $\rho_f$ is $\sim 0.12-0.2 \, \rho_0$ (where $\rho_0$
is the nuclear saturation density), the corresponding quantity
in the lattice-gas model \cite{das,cho} is $\sim 0.3-0.4 \, \rho_0$.
This wide uncertainty in both the theoretical and experimental
arena has prompted in recent times to have a closer look on the
dependence of the nuclear density with excitation \cite{sob,de}.
These calculations depend on the experimental premise that the
hot systems formed in nuclear collisions are isolated systems
having a fixed excitation energy (microcanonical ensemble);
for equilibrium, they are driven to an expanded configuration
with the maximum entropy.

The calculations of Sobotka {\it et al} \cite{sob} on nuclear 
expansion with excitation 
are schematic in nature; they employ a parametric form of 
the base density and then allow a self-similar expansion 
of the density to look for the maximum entropy configuration,
the entropy being calculated in the Fermi-gas model. In the
calculations reported in Ref. \cite{de} the same physical picture was 
followed, however, a realistic effective Hamiltonian was
used to calculate the base density profile in a Thomas-Fermi
framework. This renders the base density self-consistent 
and allows one to calculate the entropy microscopically;
furthermore, it offers the possibility of exploring the 
effect of the nuclear EoS by using a variety of effective
interactions which, however, was not reported.

In Ref. \cite{de}, calculations were performed with the 
Skyrme interaction (SkM$^*$) for only one system,
namely, $^{150}$Sm. For an expanding system pursuing 
the maximum entropy, the surface diffuseness is likely
to play a key role \cite{sob1}, a zero-range interaction 
like the Skyrme force may possibly not be suitable for studying
such a density profile. It has been reported \cite{lom} 
that a constrained expanded system in a Thomas-Fermi
approach leads to numerical instabilities  and the gradient (surface)
terms in the energy density functional were replaced with a suitable
Yukawa interaction \cite{dav}. In the present
communication, we report calculations using the modified
Seyler-Blanchard (SBM) interaction \cite{ban,de1}; it is of finite
range and momentum and density dependent. Different sets 
of SBM parameters yield different nuclear EoS.
In the present communication,  we have 
explored the role of nuclear EoS on the observables related
to an expanded mononuclear equilibrium configuration. We
have further studied the isospin and mass dependence of the
observables. In addition to expansion, a possible deformation
along the expansion path might contribute additional entropy
and mimics a fragmentation channel. We have considered volume-conserving
quadrupole deformation to explore this aspect.

\section{Model Overview}

The key tenets of the model have been reported in brief in
Ref. \cite{de}. In the following, we present a 
somewhat more comprehensive account of the theoretical framework
used in the present calculation.

\subsection{Basic Methodology}

In the experimental conditions, when two nuclei collide at intermediate
energy, a hot nuclear system with a fixed number of neutrons and
protons is formed in a nonequilibrium state. The system might also be
compressed initially resulting in a collective flow in the decompression
stage. We ignore the collective flow in the present calculation. 
As the system remains in isolation, the total excitation energy 
does not change in the subsequent evolution of the system, {\it i.e.},
the system is microcanonical. To attain equilibrium, this hot
system evolves in quest of maximum entropy. It is,
however, still possible to describe the system statistically by
an effective temperature $T$. To keep the vestiges of a canonical
temperature $T$ in the calculation has the operational 
advantage that it helps
in defining an occupation function that can be employed in evaluating
various observables like energy, entropy, etc. To obtain the equilibrium
state, we have adopted the following calculational procedure for
a given excitation energy $E^*$: 

i) The system is prepared in a
finite temperature Thomas-Fermi (FTTF) method such that the
thermal excitation energy $E_{\rm ther}(T)= E(T)-E(T=0)$ is less than 
$E^*$. Here $E(T)$ is the total energy of this system at the
temperature $T$. 

ii) Keeping the temperature fixed at $T$, the
system is allowed to undergo expansion till the total excitation
(thermal-plus-expansion) of the system equals $E^*$. The entropy for
this configuration is evaluated. 

iii) We repeat the above steps with different initial temperatures $T$
and from the plethora of different configurations so obtained, 
the one with the maximum entropy represents the equilibrium mononuclear 
configuration. 

In essence, this procedure aims at maximizing the entropy with respect
to 
the collective coordinate that describes
the expansion of the hot system under the constraint of constant
excitation energy. 
 Along the process there is a transfer from the initial thermal
energy of the system to expansion energy in search of maximal
entropy.

\subsection{Generation of the base density profile}

The FTTF framework is employed to generate the base density profile  
at the temperature $T$. The effective interaction used for the
calculation is taken to be the SBM interaction \cite{ban,de1}
which is given by
\begin{eqnarray}
v_{\rm eff}({\bf r_1},{\bf r_2},p,\rho)= -C_{l,u}\left [1-
\frac{p^2}{b^2}-d^2\left \{\rho ({\bf r_1})+\rho ({\bf r_2})\right \}^n
\right ] \frac{\exp(-r/a)}{(r/a)}.
\end{eqnarray}
 Here $r= |{\bf r_1}-{\bf r_2}|$ and $p= |{\bf p_1}-{\bf p_2}|$ are
the relative separation of the interacting nucleons in configuration
and momentum space, $\rho ({\bf r_1})$ and $\rho ({\bf r_2})$ 
are the densities at the sites of these two nucleons, and $C_l$
and $C_u$ are the strengths for like pair ($n-n$, $p-p$) and unlike
pair ($n-p$) interactions, respectively. Following
Refs.~\cite{ban,de1}, the six parameters
$C_l, C_u, a, b, d$ and $n$ are determined by reproducing the
volume energy per particle of symmetric nuclear matter, its
saturation density, volume asymmetry energy, surface energy,
the energy dependence of the real part of the nucleon-nucleus
optical potential and energies of Isoscalar Giant Monopole Resonances
(ISGMR). The density exponent $n$ governs the nuclear EoS. To
explore the effect of EoS on the relevant observables, we
have chosen two sets of SBM parameters (given in Table I)
with $n$ =1/6 and 4/3 that yield values of the nuclear incompressibility
$K_{\infty}$= 238 and 380 MeV in infinite matter, respectively. The
above interaction
with $n$ =1/6 reproduces quite well the ground-state binding
energies, root mean square charge radii,
and the ISGMR energies for a host of even-even nuclei varying
from $^{16}$O to very heavy systems. It has also been seen that
for symmetric nuclear matter, the results obtained with this
interaction agree very well \cite{uma,rud} with those calculated 
 microscopically 
with a realistic interaction in a variational
approach \cite{fri1,wir}. 

 The occupation function $n_{\tau} ({\bf r},{\bf p},T)$ for the
finite system is obtained \cite{de1} 
from the minimization of the thermodynamic
potential
\begin{eqnarray}
\Omega=E-TS-\sum_\tau \mu_\tau N_\tau .
\end{eqnarray}
It is given by
\begin{eqnarray}
n_\tau({\bf r},{\bf p},T)=\left [1+ \exp \left\{ \left(
\frac {p^2}{2m}
+V_\tau^0(r)+p^2 \, V_\tau^1(r)+V_\tau^2(r)+\delta_{\tau,Z}V_c (r)
-\mu_\tau \right) / T \right \}\right ]^{-1}.
\end{eqnarray}
In the above, $E$ and $S$ are the total energy and entropy of the
nucleus, $N_\tau$ refers to the number of neutrons or protons
($\tau$ is the isospin index), $\mu_\tau$ their chemical potentials, and
$m$ is their mass, taken to be the same for neutrons and protons. 
The quantities $V_\tau^0$ and $V_\tau^1$
are the components of the single-particle potentials corresponding to the
momentum independent and momentum dependent parts of the interaction;
$V_\tau^2$ is the rearrangement potential arising out of the density
dependent part of the interaction and $V_c$ is the Coulomb potential
which contains both the direct and the exchange parts.
Expressions for the different components of the single-particle
potential are given in Appendix-A. The momentum
dependent part $V_\tau^1$ determines the nucleon $k-$mass 
$m_{\tau,k}(r)$ as
\begin{eqnarray}
\frac{p^2}{2m_{\tau,k} (r)}=\frac{p^2}{2m}+p^2 \, V_\tau^1(r).
\end{eqnarray}
The effective single-particle (SP) potential ${\cal V}_\tau (r)$ is
given by
\begin{eqnarray}
{\cal V}_\tau (r)=V_\tau^0(r)+V_\tau^2(r)+\delta_{\tau ,Z} V_c(r).
\end{eqnarray}
The occupation function given by Eq.~(3) is then  rewritten as
\begin{eqnarray}
n_\tau({\bf r},{\bf p},T)=\left [1+ \exp \left\{ \left(
\frac {p^2}{2m_{\tau ,k}(r)} 
+{\cal V}_\tau (r)-\mu_\tau \right) / T \right\} \right ]^{-1}. 
\end{eqnarray}

The base density is obtained from
\begin{equation}
\rho_\tau (r) = \frac{2}{h^3}\int n_\tau ({\bf r},{\bf p},T)
d{\bf p} =
 A_T^*(r) \, J_{1/2}\left (\eta_\tau (r)\right ),
\end{equation}
where
\begin{eqnarray}
A_T^*(r) = \frac{4\pi}{h^3}\left [2m_{\tau ,k}(r)T\right ]^{3/2},
\end{eqnarray}
and $J_K(\eta_\tau )$ is the Fermi integral
\begin{eqnarray}
J_K(\eta_\tau )=\! \!\int_0^\infty \frac{x^K}{1+\exp(x-\eta_\tau )}dx,
\end{eqnarray}
with the fugacity $\eta_\tau$ given as
\begin{eqnarray}
\eta_\tau (r)=\left [\mu_\tau - {\cal V}_\tau (r)\right ]/T.
\end{eqnarray}
The occupation probability $n_\tau ({\bf r},{\bf p},T)$ given by 
Eq.~(6) can also be expressed as the SP occupancy in the energy
space as
\begin{eqnarray}
f(\varepsilon_\tau ,\mu_\tau ,T)=
\left [1+\exp\left \{(\varepsilon_\tau
-\mu_\tau )/T\right \}\right ]^{-1},
\end{eqnarray}
with the SP energy written as
\begin{eqnarray}
\varepsilon_\tau =\frac{p^2}{2m_{\tau ,k}(r)}+{\cal V}_\tau (r).
\end{eqnarray}
The base density from Eq.~(7) can be recast, in terms of the SP
potential and energy, as
\begin{eqnarray}
\rho_\tau (r)=\frac {1}{2\pi^2} \! \left[\frac
{2m_{\tau ,k}(r)}{\hbar^2}\right ]^{\frac {3}{2}} \!\!
\int_{{\cal V}_\tau (r)}^{\infty} \! \sqrt {\varepsilon_\tau -
{\cal V}_\tau (r)}
f(\varepsilon_\tau ,\mu_\tau ,T ) \, d\varepsilon_\tau .
\end{eqnarray}
This density profile is generated self-consistently in an iterative
procedure. The chemical potentials $\mu_\tau$ are determined
from the nucleon number conservation: 
\begin{eqnarray}
N_\tau & = &\int \rho_\tau (r) d{\bf r} \nonumber \\
& = &\int g_\tau (\varepsilon_\tau ,T)f(\varepsilon_\tau 
,\mu_\tau,T)d\varepsilon_\tau ,
\end{eqnarray}
where the single-particle level density $g_\tau (\varepsilon_\tau
,T)$ is given by
\begin{eqnarray}
g_\tau (\varepsilon_\tau ,T)  = \frac {4\sqrt 2}{\pi \hbar^3}\int
(m_{\tau ,k}(r))^{\frac {3}{2}}\sqrt {\varepsilon_\tau -
{\cal V}_{\tau}(r)}r^2 dr.
\end{eqnarray}

The modeling of a hot finite nuclear system poses some problems as 
it is thermodynamically unstable. The main difficulty arises in taking
proper account of the continuum states which are occupied at a finite
temperature as a result of which the particle density does not vanish
at large distances. The extracted observables then depend on the size of
the box in which the calculations are performed. This problem is overcome 
in the so-called subtraction procedure \cite{bonc,sur} where the hot
nucleus, assumed to be a thermalized system in equilibrium with a surrounding
gas representing evaporated nucleons, is separated from its embedding
environment. The method is based on the existence of two solutions
to the finite temperature 
Thomas-Fermi equations,
one corresponding to the liquid phase with the surrounding gas ($lg$)
and the other corresponding to the gas ($g$) phase. These two solutions
are obtained from the variational equations 
\begin{eqnarray}
\frac{\delta \Omega_{lg}}{\delta \rho_{\tau ,lg}}=0,
\end{eqnarray}
and
\begin{eqnarray}
\frac{\delta \Omega_{g}}{\delta \rho_{\tau ,g}}=0,
\end{eqnarray}
where $\Omega_{lg}$ and $\Omega_g$ are the thermodynamic potentials
[as defined in Eq.~(2)] of the respective systems with the same
chemical potentials. The base density profile for the hot nucleus
in question is given by $\rho_\tau= \rho_{\tau ,lg}-\rho_{\tau ,g}$
(which may also be called the liquid profile); this is independent of the
box size in which calculations are done. The density $\rho_\tau$
goes to zero at large distances implying vanishing surface pressure.

The expressions for $\rho_{lg}(r)$ and $\rho_g (r)$ when written in terms
of the SP potential and energy are analogous to that given in Eq.~(13).
Obviously, the nucleon $k-$mass
and the SP potentials in the two phases are different and  
\begin{eqnarray}
\int \left [\rho_{\tau ,lg}(r)-\rho_{\tau ,g}(r)\right ]d{\bf r}=N_\tau.
\end{eqnarray}

\subsection{Expansion of the system}

The density profile of an expanded
system, in principle, can be treated in a constrained
Thomas-Fermi procedure \cite{lom}. However, for large expansions, 
instability sets in which is amplified in the subtraction procedure
for the extraction of the bloated nuclear density. We, therefore,
simulate the expansion of the system through a self-similar scaling
approximation for the density: 
\begin{eqnarray}
\rho_\lambda (r)=\lambda^3\rho (\lambda r),
\end{eqnarray}
where the scaling parameter $\lambda$ is unity for the unbloated 
nucleus and decreases with expansion, lying in the range $0<\lambda \le 1$;
$\rho_\lambda (r)$ is the scaled density and $\rho (r)$ is the base
density profile generated in the FTTF framework.

Besides its simplicity, there is no {\it a priori} reason for choosing
the self-similar expansion. However, some justification can be
found on the grounds that in a simplistic situation with a harmonic
oscillator potential at small temperatures, the scaled and
the constrained density profiles are equivalent. This is shown in 
Appendix-B.

\subsection{Effect of correlations on the density}

The self-consistent 
Thomas-Fermi procedure generates
the single-particle (mean field) potential; it does not include the
coupling of the single-particle motion with the collective degrees of
freedom \cite{boh}. This coupling introduces an extra energy dependence
in the effective mass of the nucleons (in addition to the contribution
from the momentum dependence in the effective interaction). Incorporating
this effect, the total effective mass $m^*$ 
(the effective mass is position-dependent; now on, this dependence
is not explicitly shown) is then defined as
\begin{eqnarray}
m^*=m \, \frac {m_k}{m} \, \frac {m_\omega}{m},
\end{eqnarray}
where $m_\omega$ (the $\omega -$mass) is the energy or frequency
dependent
part of the effective mass. The $\omega -$mass is surface peaked and
has values generally larger \cite{has} than the nucleon mass. This
increased effective mass has the effect of bringing down the the excited
states from higher energy to lower energy near the Fermi surface,
thus increasing the many-body density of states at low excitations. The
system can then accommodate comparatively more entropy at a given
excitation energy; in the present case, in searching for the state 
with maximum entropy, these correlations might have a significant role
to play. An {\it ab initio}
determination of the $\omega -$mass \cite{has,bor} as a function of
temperature is very involved, we have therefore taken a phenomenological
form \cite{pra,shl,de2} for $m_\omega$:
\begin{eqnarray}
\frac {m_{\omega}}{m}= 1 - 0.4 \, A^{\frac {1}{3}} \exp\left
[-\left (\frac {T}{21A^{-\frac {1}{3}}}\right )^2\right ]
\frac{1}{\rho(0)} \frac{d\rho (r)}{dr} ,
\end{eqnarray}
where $\rho (0)$ is the central density of the density distribution,
the temperature $T$ is measured in MeV, and $m$ is the
nucleon mass.
The collectivity as introduced refers to the liquid phase
only; the density in the above equation is then
$\rho (r)=\rho^{lg}(r)-\rho^g(r)$ and $A$ refers to the 
liquid mass. We have thus taken the $\omega$-mass of the liquid-plus-gas
phase as $m_\omega^{lg}= m_\omega $,
the $\omega $-mass of the liquid; for the gas phase,
the frequency dependence in the effective mass is neglected,
{\it i.e.}, $m_\omega^g=m$.

A self-consistent calculation of the density profile with the inclusion
of $\omega -$mass is extremely complex; to avoid the complexities,
we have, therefore, adopted the following approach which is a realistic
extension of the method given in Ref. \cite{shl}. For each fixed
temperature $T$, we assume that the nucleus is well described by the
mean field and the effective mass given by Eq.~(20). The self-energy
of a particle in the nuclear medium is usually approximated by a local
field $(m^*/m) \, U(r)$, where $U(r)$ is a local potential (for
instance, a Woods-Saxon potential). In our case, $m_k$ is explicitly 
included in the self
consistently generated ${\cal V}(r)$. Therefore, due to the frequency
dependence in the effective mass, $g(\varepsilon )$ as defined in Eq.~(15),
is modified, taking into account the subtraction procedure, as
\begin{eqnarray}
\tilde g_\tau (\varepsilon_\tau ,T) & = &\frac 
{4\sqrt 2}{\pi \hbar^3}\int
\left [\left (m_{\tau ,k}^{lg}\frac{m_\omega}{m}\right)^{\frac {3}{2}}
\sqrt {\varepsilon_\tau -{\cal V}_\tau^{lg}(r)\frac{m}{m_\omega}} 
 \right.
 \nonumber\\
 & & \null \left.
- (m_{\tau ,k}^g)^{\frac {3}{2}}\sqrt 
{\varepsilon_\tau -{\cal V}
_\tau^g(r)}\right ]r^2 dr.
\end{eqnarray}
In Eq.~(22), the first term in the square bracket corresponds to the
liquid-plus-gas part and the second term corresponds to the gas
part. 

The chemical potential $\mu_\tau $ is now modified to 
$\tilde \mu_\tau $ to conserve the particle number:
\begin{eqnarray}
N_\tau =\int \tilde g_\tau (\varepsilon_\tau ,T)
f(\varepsilon_\tau ,\tilde \mu_\tau ,T)d\varepsilon_\tau.
\end{eqnarray}
The expressions for the densities are also accordingly modified:
\begin{eqnarray}
\tilde {\rho}_\tau^i(r)=
\frac{1}{2\pi^2\hbar^3}\left [2m_{\tau ,k}^i\frac
{m_\omega^i}{m}\right ]^{3/2}\int \sqrt{\varepsilon_\tau -
{\cal V}_\tau^i\frac{m}{m_\omega^i}}f(\varepsilon_\tau ,\tilde
\mu_\tau ,T)d\varepsilon_\tau ,
\end{eqnarray}
where $i$ refers to $lg$ or $g$.

In the present context, the following prescription is adopted
to include the effect of correlations on the scaled density
$\rho_\lambda$; the single-particle potential ${\cal V}_\lambda$
and the $k-$mass $m_{k,\lambda }$ are evaluated for the scaled 
density, the $\omega -$mass $m_{\omega ,\lambda }$ is calculated
from this density according to Eq.~(21), and then the single-particle
level density $\tilde g_\lambda$ and the chemical potential 
$\tilde \mu_\lambda $ are generated by using Eqs.~(22) and
(23) as already stated. This gives the correlated scaled density
$\tilde \rho_{\lambda ,\tau}^i (r)$.

\subsection{Energy and entropy}

In absence of correlations, the total energy of the finite unexpanded nucleus
is expressed in terms of the uncorrelated density and the occupancy
given by Eqs.~(13) and (6).
      The total energy density is obtained as
\begin{eqnarray}
{\cal E}(r)=\sum_\tau \left [{\cal E}_\tau^{\rm kin}(r)+
{\cal E}_\tau^{\rm int}(r) \right ],
\end{eqnarray}
where ${\cal E}_\tau^{\rm kin}(r)$ and ${\cal E}_\tau^{\rm int}(r)$
are the kinetic
energy and the interaction energy density, respectively. The kinetic
energy density is
\begin{eqnarray}
{\cal E}_\tau^{\rm kin}(r)& = &\frac{2}{h^3}\int \frac{p^2}{2m}n_\tau
({\bf r},{\bf p}) d{\bf p} \nonumber\\
& = & \frac{2\pi}{mh^3} \left (2m_{\tau ,k}T \right )^{5/2}
J_{3/2} \left (\eta_\tau (r) \right ),
\end{eqnarray}
where the effective mass $m_{\tau ,k}$ and the Fermi integral
$J_K (\eta)$ are defined through Eqs.~(4) and (9), respectively.

The interaction energy density is given by
\begin{eqnarray}
{\cal E}_\tau^{\rm int} (r)& = &\frac{1}{2} \left [V_\tau^0
(r)\rho_\tau (r)
+ \frac{2}{h^3} V_\tau^1(r)\int p^2 n_\tau({\bf r},{\bf p})d{\bf p} 
\right ]+{\cal E}_c(r) \delta_{\tau ,Z}\nonumber\\ 
& = &\frac{1}{2}V_\tau ^0 (r)\rho_\tau (r)+\frac{2\pi}{h^3}
\left (2m_{\tau ,k}T \right )^{5/2}J_{3/2} \left (\eta_\tau
(r) \right )V_\tau ^1(r) +{\cal E}_c(r) \delta_{\tau ,Z}.
\end{eqnarray}
The Coulomb energy density ${\cal E}_c(r)$ is the sum of the 
direct and exchange  contributions:
\begin{eqnarray}
{\cal E}_c(r)={\cal E}_c^D(r)+{\cal E}_c^{Ex}(r).
\end{eqnarray}
The direct term is obtained as
\begin{eqnarray}
{\cal E}_c^D(r)=\pi e^2\frac{\rho_p(r)}{r} \int_0^\infty \rho_p
(r^{\prime}) \left [(r+r^\prime )-|r-r^\prime | \right ]
r^\prime dr^\prime,
\end{eqnarray}
and the exchange term is calculated from the Slater approximation:
\begin{eqnarray}
{\cal E}_c^{Ex}(r)=-\frac{3e^2}{4\pi}(3\pi^2)^{1/3}\rho_p^{4/3}(r).
\end{eqnarray}
Combining Eqs.~(26) and (27), one may write (25) as
\begin{eqnarray}
{\cal E}(r)=\sum_\tau \left [\frac{2\pi}{h^3}\left (2 m_{\tau ,k}T
\right )^{5/2}J_{3/2}(\eta_\tau (r))\left (\frac{1}{m}+
V_\tau^1\right )+\frac{1}{2}V_\tau^0(r)\rho_\tau (r)+
{\cal E}_c(r)\delta_{\tau, Z}\right ].
\end{eqnarray}
The total energy is
\begin{eqnarray}
E=\int {\cal E}(r)d{\bf r}.
\end{eqnarray}

The total entropy in the Landau quasi-particle approximation for
the nuclear system with the base density profile is
\begin{eqnarray}
S=-\sum_\tau \int g_\tau (\varepsilon_\tau ,T)\left [f \ln f+(1-f)
\ln (1-f)\right ]d\varepsilon_\tau,
\end{eqnarray}
where the SP level density and occupancy are given by Eqs.~(15) and (11).

The energy of the expanded correlated system is calculated with
the suitably modified effective mass, single-particle potentials
and fugacity. 
The entropy of the correlated system can be similarly calculated.
In the subtraction procedure, the energy $E$ of the nuclear liquid 
is obtained through 
\begin{eqnarray}
E=E_{lg}-E_g,
\end{eqnarray}
where $E_{lg}$ and $E_g$ are the total energies of the liquid-plus-gas
system and of the gas alone; these energies are calculated from
Eq.~(31) with appropriate modification for expansion and correlation effects.
For the entropy, the subtracted SP level density of the form given by
Eq.~(22) 
(that includes correlation effects), but
modified for scaling is employed.

\subsection{Deformation of the system}

So far, we have taken the shape of the hot expanding system to be
spherical. In search of the maximum entropy configuration,
the system may, however, explore possible deformation paths
along with expansion mimicking fragmentation. For simplicity, 
we consider only a volume conserving quadrupole (spheroidal)
deformation at all stages of expansion and check whether there
is additional gain in entropy from deformation of the system.

In a volume conserving deformation, only the surface and Coulomb
energies change. To calculate these changes, a sharp surface
approximation to the density profile is made with 
$R_{\rm sharp}=\sqrt{\frac{5}{3} \langle r^2 \rangle }$. This also
facilitates the calculation of the entropy gain from deformation.
The excess Coulomb energy $\delta E_c(\lambda ,\beta )$ of
the expanded deformed system is given by \cite{eis}
\begin{eqnarray}
\delta E_c(\lambda ,\beta )=E_c(\lambda ,0)\left [(1-x)^{1/3}
|x|^{-1/2} F(x) -1 \right ],
\end{eqnarray}
where $\beta $, the deformation of the spheroidal system, in terms of the 
lengths $2a$ and $2c$ of its principal axes, is defined as
\begin{eqnarray}
\beta =\frac{4}{3}~\ln(c/a),
\end{eqnarray}
where $a<c$ for prolate and $a>c$ for oblate spheroids. The
constraint of volume conservation gives $a^2c=R_{\rm sharp}^3$.
The quantity $x$, in terms of $\beta $, takes the form
\begin{eqnarray}
x=\frac{3}{2}~\beta ~\frac{(1+\beta /8)}{(1+\beta /2)^2}
\end{eqnarray}
and the function $F(x)$ reads
\begin{eqnarray}
F(x)& = &\frac{1}{2}\left [\ln(1+\sqrt {x})-\ln(1-\sqrt {x})\right ],
\qquad {\rm for}~~a<c, \nonumber \\
&=& \tan^{-1}|x|^{1/2}, \qquad {\rm for}~~a>c.
\end{eqnarray}

The expression of the surface area of the deformed nucleus is
\begin{eqnarray}
{\cal A}(\lambda ,\beta )& = & 2\pi a^2+\frac {2\pi ac^2}{\sqrt {c^2-a^2}}
~\sin^{-1}\left (\frac {\sqrt {c^2-a^2}}{c}\right ), \qquad {\rm for}~~a<c, 
\nonumber \\
&=& \frac {\pi}{\sqrt {a^2-c^2}} \left [2a^2\sqrt {a^2-c^2}
+ac^2~\ln\left (\frac {a+\sqrt {a^2-c^2}}{a-\sqrt {a^2-c^2}}
\right )\right ], \qquad {\rm for}~~a>c.
\end{eqnarray}
The change in surface free energy due to deformation is given by
\begin{eqnarray}
\delta F(\lambda ,\beta )=\delta {\cal A}(\lambda ,\beta )
\, \sigma (\rho ,X,T),
\end{eqnarray}
where $\delta {\cal A} $ is the excess surface area due to deformation.
The surface tension coefficient $\sigma$ is density, temperature, and
asymmetry dependent; it is taken as
\begin{eqnarray}
\sigma (\rho ,X,T)=\sigma (\rho_0,X,T=0) \, y(\rho ) \, \alpha (T),
\end{eqnarray}
with $\sigma (\rho_0 ,X,T=0)= \sigma (\rho_0,X=0,T=0)-a_sX^2$;
here $a_s$ is the surface asymmetry coefficient taken to be 
1.7826 MeV\,fm$^{-2}$, $X=(N-Z)/A$ is the asymmetry
parameter of the nucleus, and
$\sigma (\rho_0 , X=0,T=0)=1.06$ MeV\,fm$^{-2}$ is the surface energy
coefficient of symmetric semi-infinite nuclear matter.

The surface tension has its maximum value at the saturation density
$\rho_0$. The density dependence of the surface tension coefficient
is evaluated by applying the scaling approximation on the ground-state
density profile of semi-infinite nuclear matter. 
In the sub-nuclear
density region, it is found to be well approximated by a polynomial
$y(\rho )$ of the form
\begin{eqnarray}
y(\rho )=\sum_{k=1}^4 b_k(\rho /\rho_0 )^k.
\end{eqnarray}
The values of the $b_k$ coefficients are $b_1=0.5382$, $b_2=2.3124$,
$b_3=-2.2312$, and $b_4=0.3816$. The
temperature dependence of the surface tension is taken as
\cite{rav} 
\begin{eqnarray}
\alpha (T)=\left (1+\frac {3}{2}\frac {T}{T_c}\right )
\left (1-\frac {T}{T_c} \right )^{3/2},
\end{eqnarray}
with $T_c=15$ MeV, the critical
temperature for nuclear matter.

The excess entropy from deformation is calculated as
$\delta S=-\partial (\delta F)/\partial T|_\rho$. The excess 
surface energy is obtained from 
$\delta E_{\rm surf}=\delta F + T \,\delta S$.
The total energy of the expanded deformed system can then be written as
\begin{eqnarray}
E(\lambda ,\beta ,T)=E(\lambda , 0,T)+\delta E_c(\lambda ,\beta ,T)
+\delta E_{\rm surf}(\lambda ,\beta ,T).
\end{eqnarray}
The excitation energy is given by
\begin{eqnarray}
E^*(\lambda ,\beta ,T) = E(\lambda ,\beta ,T)
- E(\lambda =1,\beta =0,T=0)
\end{eqnarray}
and the corresponding total entropy is
\begin{eqnarray}
S(\lambda ,\beta ,T)=S(\lambda ,0,T)+\delta S(\lambda ,\beta ,T).
\end{eqnarray}

\newpage

\section{Results and discussions}

To explore the density reorganization in nuclei with increasing
excitation energy, we have chosen three isobars of mass $A$=150
(namely, Cs, Sm, and Yb) and two isobars of mass $A$=40 (Ca and S)
as representative systems. The two different sets of isobars
with different isospins are chosen in order to study the mass
as well as the asymmetry dependence on the evolution of density
and some other characteristic observables related to the 
mononuclear configuration. Two sets of SBM interaction parameters (given
in Table I) with widely varying nuclear incompressibility are
used to investigate the effect of nuclear EoS on the observables
considered. All the calculations reported are performed at different
fixed excitation energies with inclusion of thermal and expansion
effects and also with incorporation of the correlation effects
introduced through the frequency dependence of the nucleon effective
mass.

The dependence of the observables related to the excited mononuclear
configuration on the nuclear EoS has not been studied so far.
In Fig.~1, the density evolution of the hot Sm nucleus 
is displayed as a function of the excitation energy
$E^*/A$ for two different EoS, one softer with $K_\infty =238$
MeV and the other one harder with $K_\infty  =380$ MeV.
Here, the quantity $\rho_c$ represents the central density of
the hot expanded nucleus in equilibrium corresponding to the
maximum entropy configuration and $\rho_0$ is the ground-state
central density evaluated in the self-consistent Thomas-Fermi
framework. Obviously, at zero excitation $\rho_c /\rho_0 =1$.
It is found that for the softer EoS, above $E^*/A ~\sim $
11 MeV, the entropy increases monotonically with expansion
and hence the mononuclear configuration ceases to exist; 
for the harder EoS this limiting excitation energy 
is somewhat higher. 
At a given excitation, the equilibrium density is seen to be lower 
for the softer EoS. This can be understood from the
following qualitative argument: the maximum entropy configuration
is an outcome of the balance between the entropy gain from
expansion and a loss due to reduction in temperature. Since for
a harder EoS, the gain in entropy from expansion is energetically
costlier, for a given excitation energy, the expansion is less
for a harder EoS. A more quantitative explanation in a simplistic
model is given in Appendix-C. The effects of correlation on the
mononuclear observables with increasing excitation energy for a
Skyrme-type (SkM$^*$) force have been discussed in an earlier
paper \cite{de}. It has been found that the changes in the 
relevant observables with the inclusion of correlation effects
with the SBM force are qualitatively the same as the SkM$^*$
interaction and therefore we do not want to delve into these
details. However, for completeness, we present some selective
results only for the nucleus $^{150}$Sm. In Fig.~1, the dashed
curve displays the results for the evolution of central density
without correlation effects for the softer SBM force. As in 
Ref. \cite{de}, the effect of correlation is found to reduce the
central density at all excitations considered.

Two sets of experimentally derived data are also shown in Fig.~1;
the filled circles are the experimental points extracted from the 
apparent level density parameters \cite{nat} for the mass selection
$140 <A<180$ and the open squares are the ones obtained from the 
Coulomb barrier systematics \cite{vio} for Au-like systems. 
It is found that the calculated results with the softer EoS
having nuclear incompressibility very close to the generally
accepted value \cite{bla} compare favorably well with those
derived from the level density parameters.

In the following, the calculations are done with
the softer EoS and the reported results are to be understood in
its terms, unless otherwise mentioned. In the upper panel
of Fig.~2, the density ($\rho_c$)
evolution with excitation for the three nuclei Cs, Sm, and Yb
is shown along with the experimental data \cite{nat,vio}.
Since the density has a profile, its characterization by a
single entity $\rho_c $ leaves some room for ambiguity. We have,
therefore, also considered an average density $\rho =A/V_{eq}$
where $V_{eq}$ is the effective volume of the excited nucleus
in equilibrium, taken as $V_{eq}=\frac{4}{3}\pi R_{sharp}^3$
where the sharp surface radius $R_{sharp}$ is calculated
as $R_{sharp}=\sqrt {\frac{5}{3}\langle r^2 \rangle }$
from the density distribution. In the lower panel 
of the figure, the density ratios
are displayed as $V_0/V_{eq}$ as a function of excitation
energy per nucleon. Here $V_0$ is the ground-state effective
volume. Inspection of the two panels  shows that the two
sets of density ratios so defined are somewhat different.
The difference is more manifest at higher excitations.
The experimental data obtained from the apparent level
density parameters lie in between the two sets of 
calculated densities.

Fig.~2  displays the asymmetry dependence of the density
evolution with excitation. It is seen that the more symmetric
nucleus has generally a higher equilibrium density at a given
excitation. This may be understood by examining the asymmetry
dependence of the incompressibility $K_A$ of finite nuclei. 
It can be written in powers of $A^{-1/3}$ with a liquid-drop
type expansion \cite{bla,tre,maj}:
\begin{eqnarray}
K_A=K_v+K_s A^{-1/3}+K_\delta \delta ^2+K_cZ^2A^{-4/3} +... ~.
\end{eqnarray}
Here $K_v$, $K_s$, $K_\delta$ and $K_c$ are the volume, surface,
asymmetry and Coulomb coefficients, respectively, and $\delta$ is
the asymmetry parameter $\delta =(N-Z)/A$. For isobars, the
asymmetry and Coulomb terms are different. From the known
\cite{tre,maj} values of $K_\delta$ and $K_c$
(which are negative), it is found that
the more symmetric nucleus has 
a larger incompressibility $K_A$  
and thus, in the
context of the discussions in relation to Fig.~1, it follows that the
equilibrium density of the more symmetric nucleus is larger. For
the nucleus with the largest charge considered here, there is a
marked deviation from the smooth decrease of density with excitation
in the energy range $6<E^*/A<8$ MeV; it is difficult to discern the exact
reason for it because of the delicate interplay of the Coulomb, thermal
and expansion effects on the density distribution. The manifestly
opposite behavior of the density ratios as defined in the two panels
of Fig.~2 can be understood by examining the density
distributions at the relevant excitations as given in Fig~3. Particularly
noticeable is the inversion of the trend of the central density
with excitation at $E^*/A =7$ MeV. The steep decrease of the density
ratio given by $V_0/V_{eq}$ can be understood from the rapid
growth of the density tail with excitation. The evolution of density 
with excitation energy for the $A=~40$ systems (Ca and S) is displayed
in Fig.~4. As in the case of heavier systems explored (Fig.~2), the density
for the more symmetric system is higher; the asymmetry effect on
the density is seen to be more prominent for the heavier systems.
It is further found that for the same asymmetry, the heavier
systems have somewhat higher equilibrium density at a given excitation.
It is noted that the two definitions used for the characterization
of density yield values which are not much different for the 
lighter systems.

The expansion of the nucleus has an important bearing on the correlation
of excitation energy with temperature. The correlation (caloric curve)
so obtained for the expanded hot nucleus $^{150}$Sm is displayed
in Fig.~5. At a fixed excitation energy, the system cools down
with expansion, and therefore the recorded temperature at the
equilibrium configuration is significantly lower than that for
the unexpanded nucleus prepared initially with the same excitation.
In all of our calculations, the temperature refers to the canonical
temperature. We have checked that the microcanonical temperature
obtained from $T^{-1}=~\partial S_{eq}/\partial E^*$, 
is not much different from the canonical one. To explore the role
of the EoS on the correlation between $E^*$ and $T$,
the caloric curves obtained from both the softer ($K_\infty =238$ MeV)
and the harder ($K_\infty =380$ MeV) EoS are shown in Fig.~5.
It is seen that at a given $E^*$, the equilibrium temperature
is significantly higher, particularly at high excitations, for the
stiffer EoS. Tentatively, this can be understood from the fact
that the equilibrium density is higher for a harder EoS (Fig.~1),
the corresponding expansion is less and so more energy is locked
in the thermal mode with a resultant higher temperature.
The same conclusion is arrived at from the simplistic Fermi-gas
type framework presented in Appendix-C. A representative set
of experimental data \cite{cib} for medium-mass nuclei
are also shown in the figure. It is seen that the calculated
caloric curve with the softer EoS compares very well with the 
experimental data. The dashed curve represents the caloric curve
without inclusion of correlation effects ($m_\omega /m=1$).
As in Ref. \cite{de}, introduction of correlation is seen to
reduce the temperature for the equilibrium configuration at all
excitation energies considered.

 The dependence of the nuclear caloric curve on mass and 
isospin asymmetry is shown in Fig.~6. The top panel refers
to the heavier isobars and the bottom panel to the lighter ones.
It is apparent from the figures that the caloric curves are
nearly independent of asymmetry at lower excitations. For
$E^*/A~ >$ 5 MeV, the isospin dependence becomes increasingly
prominent with increasing excitation, the equilibrium temperature
is higher for the more symmetric system. 
The mass dependence on the caloric curve is seen from an
examination of the results for pairs of nuclei which have almost
the same asymmetry like Sm and S or Yb and Ca. 
It is found that the equilibrium temperature for the lighter nuclei
are comparatively always lower at all excitations. This
behavior of the caloric curve with mass and isospin can be
explained from the fact [using Eq.~(47)] that for the same
asymmetry, heavier nuclei have larger incompressibility;
also, for isobars, the ones with more asymmetry have an effectively
lower incompressibility.

The caloric curves show an interesting feature, namely, the
occurrence of negative heat capacity, generally beyond an
excitation energy $E^*/A \sim 8$ MeV. Intuitively, one
understands that if a system with a given excitation expands,
it does so at the cost of thermal energy and hence there may
be a density region when temperature may decrease with increasing 
excitation if the system expands much in pursuit of maximum 
entropy. In a simplistic model in Appendix-C, we show that this
occurs in the density region $1/4 < \rho/\rho_0 < 5/8$. It is also
found that the more symmetric systems have lesser bends in the
caloric curves (for Yb, in the excitation energy range explored,
the bend shows up only later at $E^*/A~ \sim $11 MeV), 
this is because they offer more resistance towards volume 
expansion as already explained. 

The expansion energy $E_{\rm expn}$ constitutes a part of the total
excitation. It is defined as 
$E_{\rm expn}=~E(\lambda ,T) - E(\lambda =1,T)$.
In Fig.~7, it is displayed as a function of $E^*$ for the different
systems. In the bottom panel, the effect of the EoS on the expansion energy
is shown for the nucleus $^{150}$Sm. It is seen that $E_{\rm expn}$
decreases with stiffness of the EoS; the same is true for all
other systems studied. The filled circle refers to an experimental
estimate \cite{vio} for medium-mass nuclei ($A \sim
160-180$). This is very
close to our calculations with the softer EoS for the $^{150}$Sm
nucleus. The top and middle panels refer to calculated results
for isobars with $A$=150 and 40, respectively. As already stated,
the more symmetric nuclei have effectively a harder EoS, this is 
reflected in the comparatively lower expansion energy for these nuclei. 
It is seen that in general, the expansion energy comprises a
significant part of the total excitation.

The state with maximum entropy $S(\lambda_{eq},T_{eq})$ corresponds
to the state with maximum probability. Other configurations
with different values of the scale parameter $\lambda $ and
temperature
$T$ are also probable. Their probability is determined by the
entropy profile $S(\lambda ,T)$ at a fixed excitation energy:
\begin{eqnarray}
W(\lambda ,T) \propto~ e^{S(\lambda ,T)}.
\end{eqnarray}
The $n$th moment of the central density $\rho_c$ is then
\begin{eqnarray}
\langle \rho_c^n(\lambda ,T) \rangle=~\frac{\int e^{S(\lambda ,T)}
\rho_c^n(\lambda ,T)d\rho}{\int e^{S(\lambda ,T)}d\rho} ~.
\end{eqnarray}
Similarly, for the temperature $T$ one has
\begin{eqnarray}
\langle T^n(\lambda ,T) \rangle=~\frac{\int e^{S(\lambda ,T)}
T^n(\lambda ,T)dT}{\int e^{S(\lambda ,T)}dT} ~.
\end{eqnarray}
Equations (49) and (50) allow one to calculate the mean and the variance
of $\rho_c$ and $T$ at a fixed excitation. Figs.~8 and 9 display
the entropy profile for the systems $^{150}$Cs and $^{150}$Yb at
different fixed excitations $E^*/A = 4$, 6, 8, and 10 MeV
as a function of the central density (Fig.~8)
and temperature (Fig.~9)
of the expanded configurations. The arrows in
the figures indicate the configurations with the maximum
probabilities, the full ones refer to Yb and the broken ones
correspond to Cs. The entropy profiles
are seen to flatten with excitation. This implies larger fluctuations
at higher excitations.

For a thermodynamic system, the average and the most probable
(equilibrium) value of an observable are the same. For a finite
system, however, they may differ. Experimentally, it is the
average value that one measures. The average temperatures
for the $A$=150 isobaric systems are displayed in
panel (a) of Fig.~10; the corresponding quantities for 
$^{40}$Ca and $^{40}$S are shown in panel (c). 
 Panels (b) and (d) display the variances 
in temperature for the isobaric systems with
$A$=150 and 40, respectively. The equilibrium
temperatures as shown in Fig.~6 are somewhat larger than the
corresponding average values.
Fluctuations build up with 
excitation. At a given excitation energy, the fluctuations
for the lighter systems are comparatively higher as expected.
For the heavier systems, a sudden increase in fluctuation
is observed beyond $E^*/A \sim$ 9 MeV, whereas the increase
is relatively smooth for the lighter systems. In Fig.~11, the
average values of the specific volume
$v_c$ (=$1/\rho_c$) in units of $v_0$ ($= 1/\rho_0$)
and their variances for the systems considered are shown. 
The equilibrium volume is found to be somewhat 
less than the average value. As in the case of
temperature, the fluctuations in volume rise smoothly up to
an excitation energy $E^*/A \sim$ 9 MeV, beyond which the
build-up is very sudden. This sudden build-up is more
pronounced in comparison to that in temperature for both
light and heavy systems. This large density fluctuation 
indicates that beyond $E^*/A \sim$ 9 MeV, the systems become
unstable and break up into many pieces. It turns out
that the negative branch of the heat capacity and the onset
of large fluctuations occur at around the same excitation   
energy indicating a possible close correlation between them.

Along with expansion, the excited nucleus may undergo deformation
if that is profitable from the entropy considerations. From the 
interplay of Coulomb and surface energies, a barrier is found
along the deformation path; the barrier decreases with increasing
temperature and decreasing density (expansion). 
At a particular temperature, for the deforming system, in our
calculations the scale parameter $\lambda $ is taken
such that the total excitation at the top of the barrier matches the
given excitation $E^*/A$. This is repeated for different temperatures
and the maximum entropy among these different configurations is
selected. If this entropy exceeds that for the expanded spherical
equilibrium configuration, then deformation is favored leading
to the fragmentation channel.

The extra entropy $\Delta S$ gained from deformation over that at
spherical equilibrium shape is displayed in the upper panels of
Fig.~12 for the nuclei Sm and Cs as a function of excitation energy.
As mentioned, we have considered a volume-conserving quadrupole
deformation. For $E^*/A$ less than $\sim $ 4 MeV, $\Delta S$ is negative
in the restricted deformation space chosen; it is thus seen that 
only above this excitation domain, fragmentation resulting
from deformation is more favorable. With deformation, the
maximum entropy configuration occurs at a lower temperature
at a given excitation; the corresponding caloric curves are shown 
in the bottom panels. In all the calculations, it is found that
the oblate shapes are inhibited due to their lower entropy.

\section{Concluding remarks}

In the present work, the gross features of an expanding mononuclear
configuration at moderate and higher excitations have been dealt with
in a semi-microscopic framework. A finite range, momentum and 
density dependent realistic effective interaction has been employed
for our investigation. Renormalization of the nucleon mass coming
from the coupling of the single-particle motion with the surface
degrees of freedom is seen to play an important role in the correlation
of entropy with excitation in  a microcanonical formulation, this
has been taken into account phenomenologically to avoid complexity.
The density dependence in the interaction leaves room to vary
the nuclear equation of state well within the ambits of accepted
nuclear parameters. We have explored the influence of the EoS
on mononuclear observables and find that at a fixed excitation,
a nucleus with a softer EoS has more expansion, the corresponding
expansion energy is found to be larger and thus its effective
temperature is comparatively lower. 

The resultant mononuclear caloric curve
and equilibrium density compare very well with the experimental 
data with the chosen softer EoS with nuclear incompressibility
$K_\infty$~=238 MeV, which is close to its well-accepted value.
We have further studied the mass and isospin dependence on the
mononuclear observables. For nuclei with same asymmetry but
higher mass, it is found that the system expands less and
consequently has a higher temperature; this is also true for 
nuclei with same mass but less asymmetry. All these observations 
could be explained from the fact that heavier and symmetric systems
have effectively higher incompressibility.

In the experimentally constructed nuclear caloric curves \cite{poc,cib},
a plateau is observed in the excitation energy range $\sim 3-8$ MeV/$A$.
This is indicative of infinite specific heat. In some analyses, 
a negative specific heat \cite{dago} in this energy domain is also
suggested. These are pointers to a possible liquid-gas phase
coexistence. In our calculations, the plateau comes naturally
from nuclear expansion, so also its bending (negative specific heat)
at a relatively higher excitation. The fluctuations in temperature
and volume build up with excitation. The rapid growth of these
fluctuations at $E^*/A \sim$ 9 MeV is suggestive of the instability
of the mononuclear configuration against prompt multifragmentation.
The occurrence of the negative heat capacity and the sudden growth
of fluctuations at around the same excitation energy indicate a 
close correlation between them.

While expanding, the nuclear system may gain further entropy from
fragmentation into a number of pieces, mostly from the generation
of new surfaces. We have dealt with this in a simplistic way through
a volume-conserving quadrupole deformation as a precursor to
fragmentation. It is found that above $E^*/A \sim$ 4 MeV, the
systems generally favor deformation. The evolution of the mononuclear
configuration into a number of fragments may be treated through
higher multipole deformation; it would be an involved task, but
it is worth investigating.

\acknowledgments{S.K.S. and J.N.D. acknowledge the financial support
from CSIR and DST, Government of India, respectively. M.C. and X.V.
acknowledge financial support from Grants No.\ FIS2005-03142 from MEC
(Spain) and FEDER, and No.\ 2005SGR-00343 from Generalitat de
Catalunya.}

\appendix

\begin{center}
{\bf Appendix-A\\
The single-particle potentials}
\end{center}

 \renewcommand{\theequation}{A-\arabic{equation}}
\setcounter{equation}{0}
The components of the single-particle potentials for the
SBM interaction (1) entering in
Eq.~(3) can be written as \cite{de1},
\begin{eqnarray}
V_\tau^0(r_1)& = &-2\pi a^2\int_0^\infty \left [1-d^2\left \{\rho (r_1)
+\rho (r_2)\right \}^n \right ]u(r_1,r_2) \left [C_l \rho_\tau (r_2)
+C_u \rho_{-\tau} (r_2) \right ]r_2^2 dr_2
\nonumber\\[2mm]
& & \null
+\frac{a^2}{\pi b^2 \hbar^3} \int_0^\infty u(r_1,r_2)
\left [C_l \left (2m_{\tau ,k}(r_2)T \right )^{5/2} J_{3/2} \left (
\eta_\tau (r_2) \right )
 \right.
 \nonumber\\[2mm]
 & & \left. \null
+ C_u\left (2m_{-\tau ,k}(r_2)T\right )^{5/2}
J_{3/2}\left (\eta_{-\tau}(r_2)\right ) \right ] r_2^2 dr_2,
\end{eqnarray}
\begin{eqnarray}
V_\tau ^1(r_1) & =& \frac{2\pi a^2}{b^2}\int_0^\infty u(r_1,r_2)
\left [C_l\rho_\tau (r_2)+C_u\rho_{-\tau} (r_2) \right ] r_2^2 dr_2 ,
\end{eqnarray}
\begin{eqnarray}
V_\tau ^2 (r_1)& = &2\pi n a^2 d^2 \sum_{\tau ^\prime } \{ \left
[C_l\rho_{\tau ^\prime }(r_1)+C_u\rho_{-\tau ^\prime} (r_1)\right ]
 \nonumber\\[2mm]
 & &  \times
\int_0^\infty \rho_{\tau ^\prime}(r_2)u(r_1,r_2)
\left [\rho (r_1)+\rho (r_2) \right ]^{n-1} r_2^2 dr_2 \},
\end{eqnarray}
with
\begin{equation}
\rho(r) = \rho_\tau (r) + \rho_{-\tau} (r) 
\quad {\rm and} \quad
u(r_1,r_2)=\left [e^{-|r_1-r_2|/a}-e^{-(r_1+r_2)/a} \right ]/(r_1r_2).
\end{equation}
In these equations, if $\tau $ refers to proton, $-\tau $ refers to
neutron and vice versa. From the structure of Eq.~(A-3),
it is seen that the 
rearrangement potential $V_\tau ^2 (r)$ is isospin independent.

 The Coulomb single-particle potential is the sum of the direct and
exchange contributions:
\begin{eqnarray}
V_c(r_1)=V_c^D(r_1)+V_c^{Ex}(r_1),
\end{eqnarray}
where 
\begin{eqnarray}
V_c^D(r_1)=\frac{2\pi e^2}{r_1}\int_0^\infty \left [(r_1+r_2)-|r_1-r_2|
\right ] \rho_p(r_2)r_2dr_2,
\end{eqnarray}
and
\begin{eqnarray}
V_c^{Ex}(r_1)=-e^2(3/\pi )^{1/3}\rho_p^{1/3}(r_1).
\end{eqnarray}

\appendix

\begin{center}
{\bf Appendix-B\\
Equivalence of scaled and constrained density}
\end{center}

 \renewcommand{\theequation}{B-\arabic{equation}}
\setcounter{equation}{0}

In this Appendix we show the equivalence between the scaled density
and the constrained density for a harmonic oscillator (HO) potential.

From Eqs.~(7) and (8), the density at a finite temperature for a nucleus
(ignoring the isospin) can be written in the form
\begin{eqnarray}
\rho (r)=\frac {8\pi}{h^3}(2mT)^{3/2}J_{1/2}(\eta ),
\end{eqnarray}
where $\eta =(\mu -V)/T$, $V$ being the single-particle 
potential. For small temperatures, $\eta$ is large.
When $\mu < V$, $\eta$ is large negative and then $J_{1/2}(\eta)$
$\sim e^{\eta}$ leading to vanishing density. When $\mu > V$, $\eta$
is large positive and then $J_{1/2}(\eta) \simeq \frac{2}{3}\eta^{3/2}$,
and the density is given by
\begin{eqnarray}
\rho (r)=\frac {2}{3 \pi^2} \left (\frac {2m}{\hbar^2}\right )^{3/2}
(\mu -V)^{3/2}.
\end{eqnarray}
For a harmonic oscillator potential with oscillator frequency $\omega $, 
$V=\frac {1}{2}m\omega^2r^2$,
the corresponding density is
\begin{eqnarray}
\rho (r)=B \left(\mu -\frac {1}{2}m\omega^2r^2\right)^{3/2},
\end{eqnarray}
where $B=\frac {2}{3\pi^2}\left (\frac {2m}{\hbar^2}\right )^{3/2}$.
The total number of particles for this density is
\begin{equation}
A = \int \rho (r)d{\bf r} 
= \frac {2}{3}\left (\frac {\mu}{\hbar \omega} \right )^3.
\end{equation}

For the density given by Eq.~(B-3), following Eq.~(19),the scaled density is
\begin{eqnarray}
\rho_\lambda (r)=B\left [\mu\lambda^2-\frac{1}{2}m(\omega
\lambda^2)^2r^2\right ]^{3/2}.
\end{eqnarray}
It is seen that the chemical potential and the 
oscillator frequency are now scaled as
\begin{eqnarray}
\mu_\lambda \rightarrow \mu\lambda^2 \qquad {\rm and}
\qquad \omega_\lambda \rightarrow \omega\lambda^2.
\end{eqnarray}
The constrained Hamiltonian density constraining the particle number
and the rms radius is given by
\begin{eqnarray}
{\cal H}(r)={\cal E}^{\rm kin}(r)+(V-\alpha r^2)\rho_\alpha (r)-
\mu_\alpha \rho_\alpha (r),
\end{eqnarray}
where ${\cal E}^{\rm kin}$ is the kinetic energy density,
$\mu_\alpha$ constrains the
total number of particles and $\alpha$ constrains the rms radius
to given values $A$ and $R_\alpha$, respectively, so that
\begin{eqnarray}
A=\int \rho_\alpha (r)d{\bf r},
\end{eqnarray}
and
\begin{eqnarray}
R_\alpha ^2=\frac {1}{A}\int r^2\rho_\alpha (r)d{\bf r}.
\end{eqnarray}
In the absence of the constraint on the rms radius, 
$\rho_\alpha \rightarrow \rho $ and $R_\alpha \rightarrow R$, the
unconstrained radius.

In analogy to Eq.~(B-2), at low temperatures, 
the density corresponding to the Hamiltonian given
by Eq.~(B-7), for a HO potential is written as
\begin{eqnarray}
\rho_\alpha (r)=B \left [\mu_\alpha -\frac {1}{2}m\omega^2r^2
+\alpha r^2\right ]^{3/2},
\end{eqnarray}
which can be recast as
\begin{eqnarray}
\rho_\alpha (r)=B\left [\mu_\alpha -\frac {1}{2}m\omega_\alpha^2r^2
\right ]^{3/2}, 
\end{eqnarray}
where $\omega_\alpha^2=\omega^2- 2\alpha / m$. Again, following
Eq.~(B-4), the total particle number from Eq.~(B-8) is given by
\begin{eqnarray}
A=\frac {2}{3}\left (\frac {\mu_\alpha}{\hbar\omega_\alpha} \right )^3.
\end{eqnarray}
From Eqs.~(B-9) and~(B-12), it then follows that
\begin{eqnarray}
\hbar\omega_\alpha R_\alpha ^2=\frac {\hbar^2}{2m}\frac {3}{2}
\left (\frac {3A}{2}\right )^{1/3}.
\end{eqnarray}
This is also true for the unconstrained density ($\alpha$ =0):
\begin{eqnarray}
\hbar\omega R^2=\frac {\hbar^2}{2m}\frac {3}{2}\left (\frac {3A}{2}
\right )^{1/3}.
\end{eqnarray}
From Eqs.~(B-13) and~(B-14), 
\begin{eqnarray}
\hbar\omega_\alpha =\frac {R^2}{R_\alpha^2}\hbar\omega.
\end{eqnarray}
Similarly, from Eqs.~(B-12) and~(B-15),
\begin{eqnarray}
\mu_\alpha =\mu \frac{R^2}{R_\alpha^2}.
\end{eqnarray}
Therefore, the constrained density given by Eq.~(B-11) is rewritten as
\begin{eqnarray}
\rho_\alpha (r)=B\left [\mu \frac{R^2}{R_\alpha^2} -\frac{1}{2}
m\left (\omega \frac{R^2}{R_\alpha^2}\right )^2 r^2\right ]^{3/2}.
\end{eqnarray}
Identifying $\lambda$ as $R/R_\alpha$, one can see the equivalence of
the scaled density given by Eq.~(B-5) with the constrained density.

\appendix

\begin{center}
{\bf Appendix-C\\
EoS dependence of equilibrium density and temperature}
\end{center}

 \renewcommand{\theequation}{C-\arabic{equation}}
\setcounter{equation}{0}

In this Appendix, we show in a simplistic model why the 
density and temperature at equilibrium are higher for a 
stiffer EoS at a given excitation $E^*$.

The expansion energy for the decrease of density from $\rho_0$
to $\rho $ is
\begin{eqnarray}
E_{\rm expn}~=~\frac{K_A}{18}(1-\rho /\rho_0)^2.
\end{eqnarray}
This is of the same form as suggested by Friedman \cite{frie};
$K_A$ is the incompressibility of the nucleus.

The thermal energy of a hot nucleus, in a Fermi-gas type
model is
\begin{eqnarray}
E_{\rm ther}~=~a \, T^2~,
\end{eqnarray}
where the density dependence of the level density parameter is
given as
\begin{eqnarray}
a~=~\frac{\pi ^2}{4\varepsilon_F(\rho )}.
\end{eqnarray}
The latter can be written in the form
\begin{eqnarray}
a~=~b\rho ^{-2/3},
\end{eqnarray}
where $b$ is a constant. The total excitation energy is given by
the sum of the expansion and the thermal energy,
\begin{eqnarray}
E^*~=~E_{\rm expn}+E_{\rm ther}~.
\end{eqnarray}

The entropy is given by $S(\rho ,T)=~2aT=~2b\rho ^{-2/3}T$.
At equilibrium, $S$ is a maximum. Because of the constraint 
that the sum of the thermal and expansion energies is equal
to the given $E^*$, the quantity to be maximized is
\begin{eqnarray}
\tilde {S}(\rho ,T)=~2b\rho ^{-2/3}T-\frac{1}{\nu }
\left[b\rho ^{-2/3}T^2+\frac{K_A}{18}(1-\rho/\rho_0)^2-E^* \right ],
\end{eqnarray}
where $\nu $ is a Lagrange multiplier. Then $\partial \tilde {S}/
\partial T =0$ yields $\nu =T$, and $\partial \tilde {S}/\partial
\rho=0$ leads to
\begin{eqnarray}
bT^2=~\frac {K_A}{6\rho_0}\rho ^{5/3}(1-\rho /\rho_0),
\end{eqnarray}
resulting in
\begin{eqnarray}
E_{\rm ther}=\frac {K_A}{6}\frac {\rho}{\rho_0}(1-\rho /\rho_0).
\end{eqnarray}
From Eqs.~(C-1) and (C-8), the total excitation is
\begin{eqnarray}
E^*=~\frac {K_A}{18}\left [1+\frac {\rho }{\rho_0}-2\left(\frac {
\rho }{\rho_0}\right)^2 \right ].
\end{eqnarray}
Thus at a fixed $E^*$, if $K_A$ is higher, the quantity
in the square bracket should be smaller. This is true
if $\rho /\rho_0$ increases with $K_A$, 
in the simplistic model this is satisfied only in the
range $1/4<\rho /\rho_0<1$, which, however, covers 
almost the whole density range of our concern. From Eq.~(C-7),
\begin{eqnarray}
\frac {dT}{dE^*}=~\frac {K_A \rho ^{2/3}}{36 \rho_0 \, b \, T}
\left [ \left( 5-8\frac {\rho }{\rho_0} \right)
\frac {d\rho}{dE^*} \right ].
\end{eqnarray}
Eq.~(C-10) shows that $dT/dE^*$ is negative if the quantity in 
the square bracket is negative. From Eq.~(C-9), one finds that
$d\rho /dE^*$ is negative in the density range $1/4<\rho /\rho_0
<1$. In this schematic model, one thus finds that for
$1/4<\rho /\rho_0<5/8$, the temperature decreases with excitation.

\newpage

\vskip 2cm

TABLE I. The parameters of the effective interactions used in the
present paper (in MeV\,fm units) and their bulk incompressibility
modulus (in MeV).

\vskip 1cm

\begin{ruledtabular}
\begin{tabular}{ccccccc}
{$n$} & {$C_l$} & {$C_u$} & {$a$} & {$b$} & {$d$} & {$K_{\infty}$} \\
\hline
1/6 & 291.7 & 910.6 & 0.6199    & 928.2 & 0.879 & 238 \\
4/3 & 210.7 & 653.2 & 0.529\;\; & 615.4 & 0.924 & 380
\end{tabular}
\end{ruledtabular}

\newpage

\centerline
{\bf Figure Captions}
\begin{itemize}
\item[Fig.\ 1] The calculated equilibrium density of $^{150}$Sm as a
function of excitation energy per nucleon. The thick line corresponds 
to the calculations with the soft EoS ($K_\infty$=238 MeV), 
and the thin line refers to those with the hard EoS ($K_\infty$=380
MeV). The dashed line refers to calculations without correlation 
effects for the softer EoS.
The filled circles are the experimental data from Ref. \cite{nat};
the open squares are the ones from Ref. \cite{vio}.
 
\item[Fig.\ 2] The calculated equilibrium densities as a function
of excitation energy for $A=150$ isobars in the two definitions of
density as explained in the text. The experimental points are the
same as in Fig.~1.  

\item[Fig.\ 3] The equilibrium proton density profile for the system
$^{150}$Yb at excitation energies of 6, 7, and 8 MeV per nucleon.

\item[Fig.\ 4] Same as in Fig.~2 for the nuclei $^{40}$Ca and
$^{40}$S.

\item[Fig.\ 5] The mononuclear caloric curve for the system $^{150}$Sm.
The thick and thin lines refer to soft and hard EoS, respectively.
The dashed line corresponds to the caloric curve calculated without
correlation effects  
{\bf for the softer EoS.}
The experimental
data points (filled circles) are from Ref. \cite{cib}.

\item[Fig.\ 6] Mononuclear caloric curves showing isospin effects for
$A=150$ isobars (top panel) and $A=40$ isobars (bottom panel). The
experimental data are as in Fig.~5.
 Information on the mass dependence of the caloric curves can be
gained by comparison of the curves of the pair of nuclei $^{150}$Sm
and $^{40}$S that have asymmetry $X\sim 0.2$, with those of the nuclei
$^{40}$Ca and $^{150}$Yb that are symmetric or nearly symmetric.

\item[Fig.\ 7] Expansion energy as a function of total excitation
energy
for $A=150$ isobars (top panel) and $A=40$ isobars (middle panel). The
EoS dependence of the expansion energy for $^{150}$Sm is illustrated
in the bottom panel. The filled circle corresponds to an experimental
estimate \cite{vio}.

\item[Fig.\ 8] Entropy profile at different fixed excitations as a
function of the central density for the nuclei $^{150}$Yb and
$^{150}$Cs. The equilibrium configurations are marked by arrows, full
ones for Yb and the broken ones for Cs.

\item[Fig.\ 9] Entropy profile at different fixed excitations as
a function of temperature for the systems $^{150}$Yb and $^{150}$Cs.
The arrows have the same meaning as in Fig.~8.

\item[Fig.\ 10]  The average values of temperature as a function
of excitation energy are shown in panels (a) and (c) for
$A=150$ and $A=40$ isobars, respectively. In panels (b) and (d),
the corresponding variances in temperature are shown.

\item[Fig.\ 11] The average values of the specific volumes and their
variances for the different systems considered are displayed.

\item[Fig.\ 12] The entropy gain $\Delta S$ from deformation as a
function of $E^*/A$ for the systems $^{150}$Sm and $^{150}$Cs are
displayed in the upper panels. The caloric curves with (full line)
and without (dashed line) deformation are shown in the lower panels.

\end{itemize}

\newpage

\vspace*{-1cm}
\begin{figure}[t]
\includegraphics[width=0.65\textwidth,angle=270,clip=false]{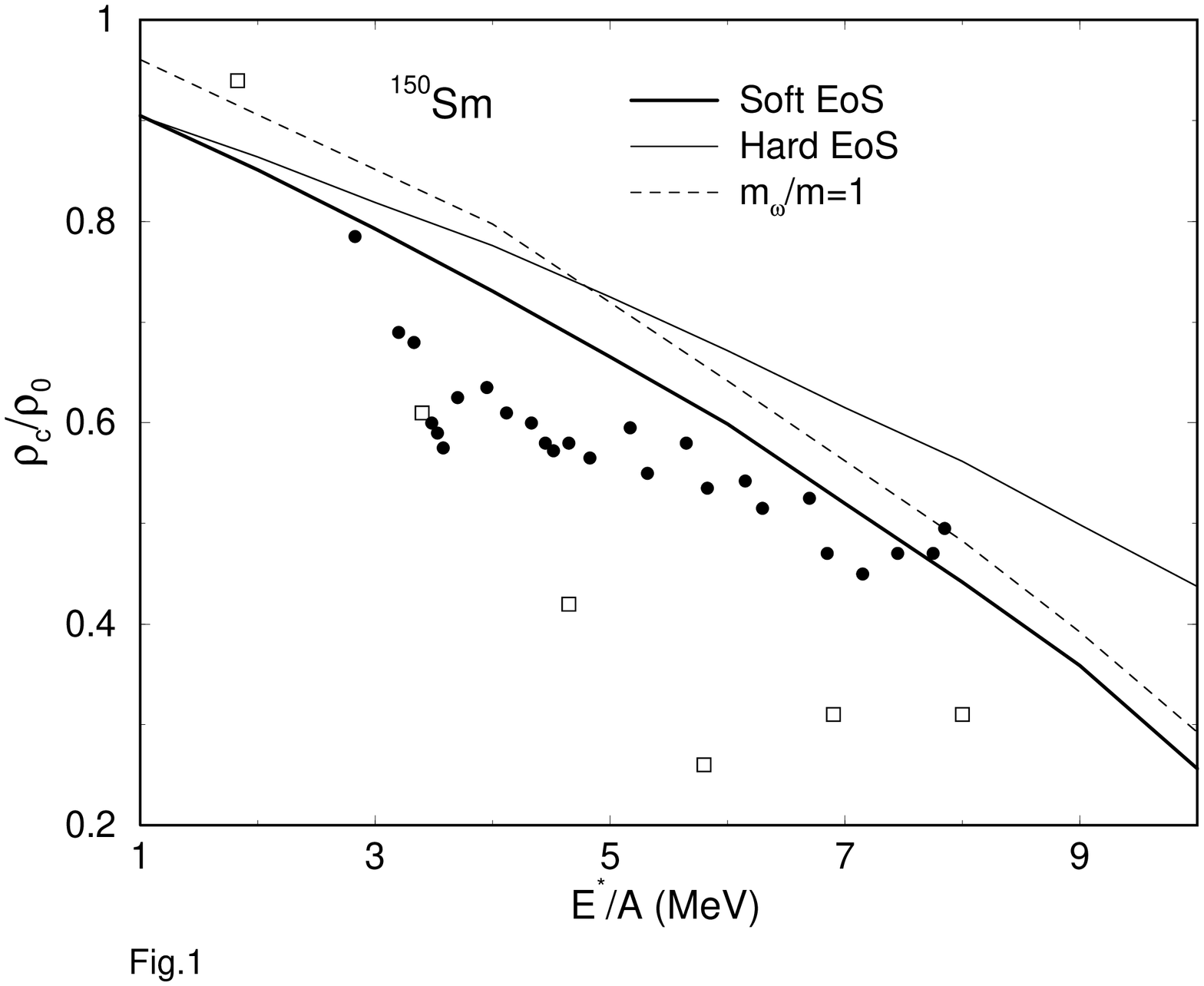}
\end{figure}

\begin{figure}[b]
\includegraphics[width=0.65\textwidth,angle=270,clip=false]{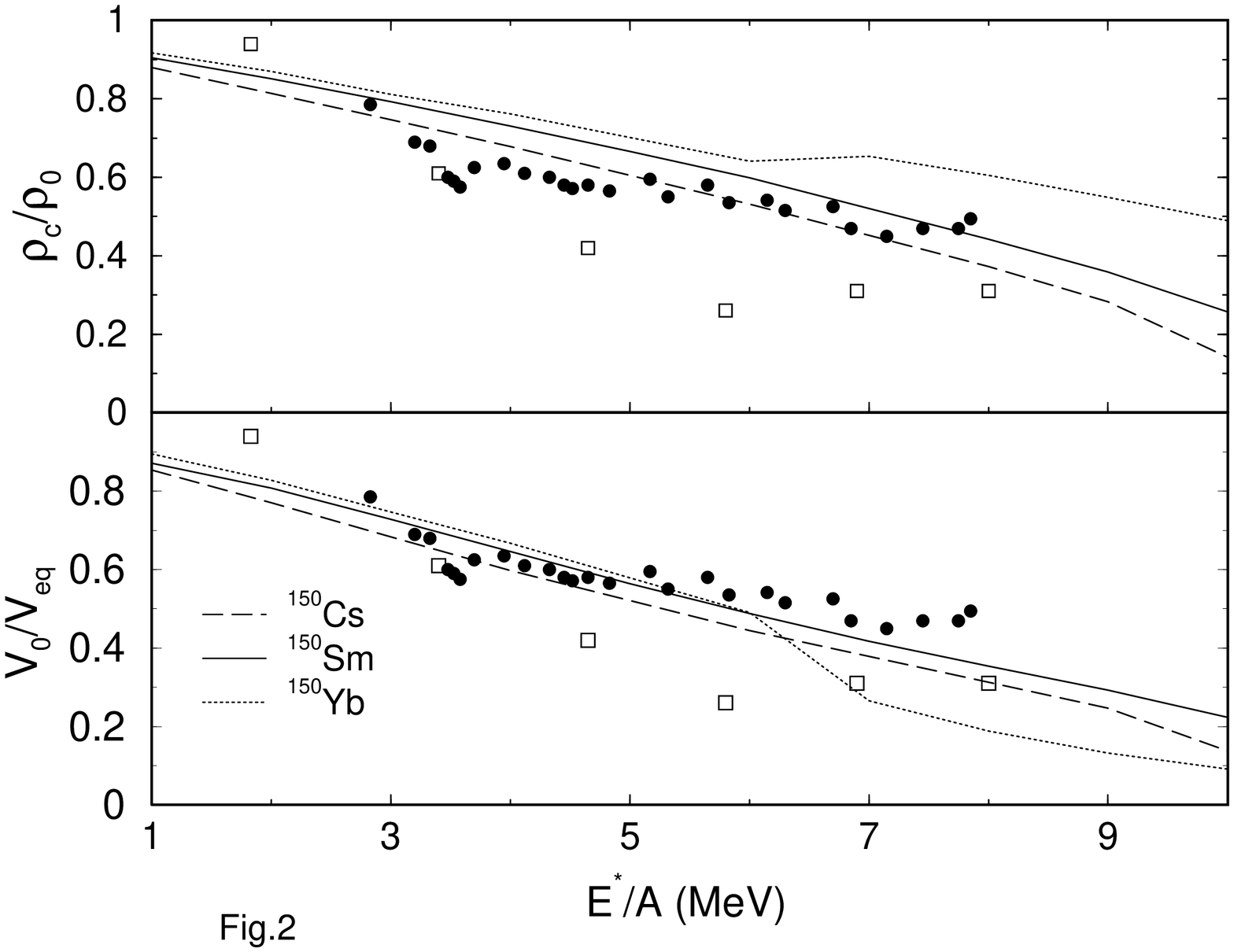}
\end{figure}

\vspace*{-1cm}
\begin{figure}[t]
\includegraphics[width=0.65\textwidth,angle=270,clip=false]{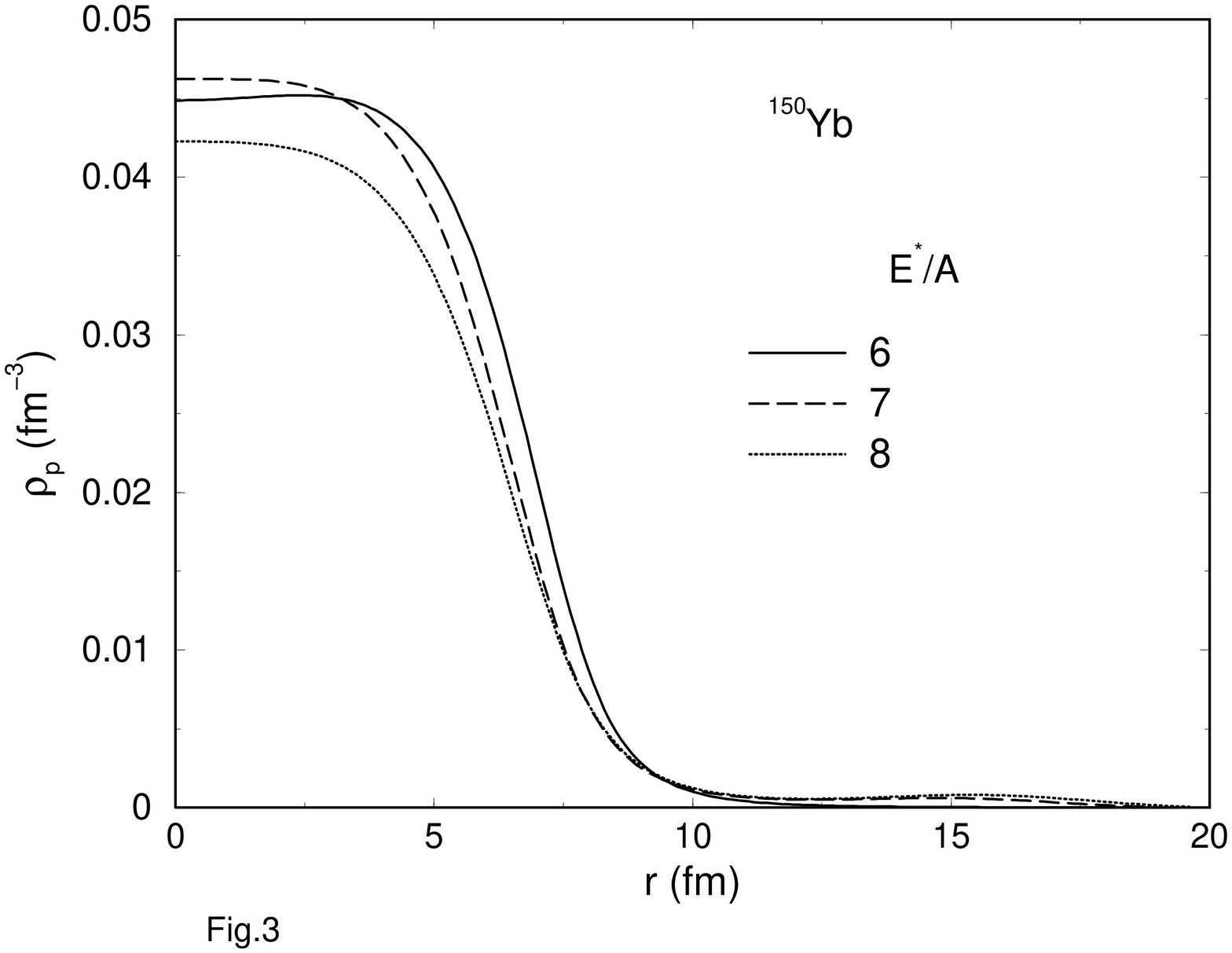}
\end{figure}

\begin{figure}[b]
\includegraphics[width=0.65\textwidth,angle=270,clip=false]{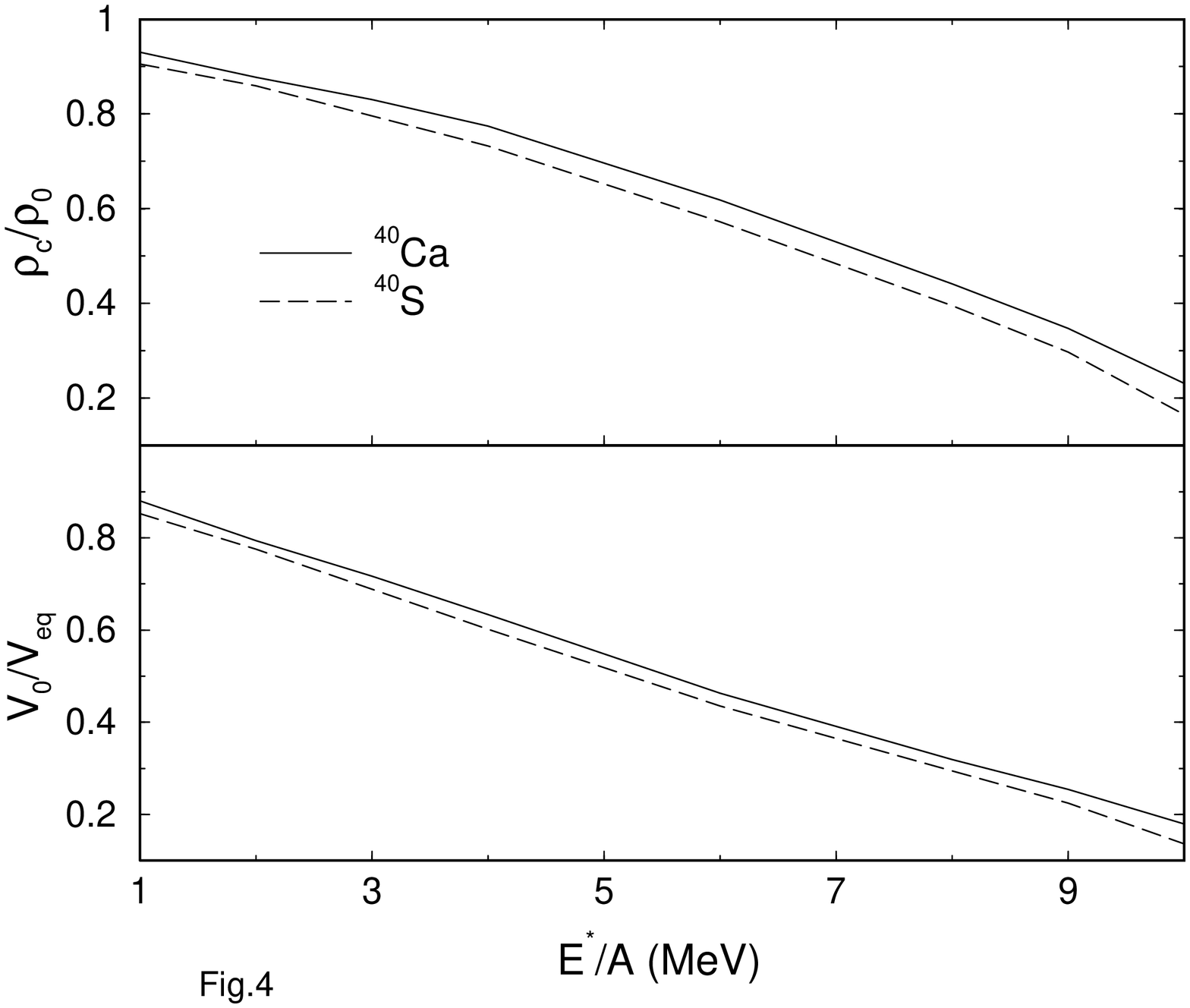}
\end{figure}

\vspace*{-1cm}
\begin{figure}[t]
\includegraphics[width=0.65\textwidth,angle=270,clip=false]{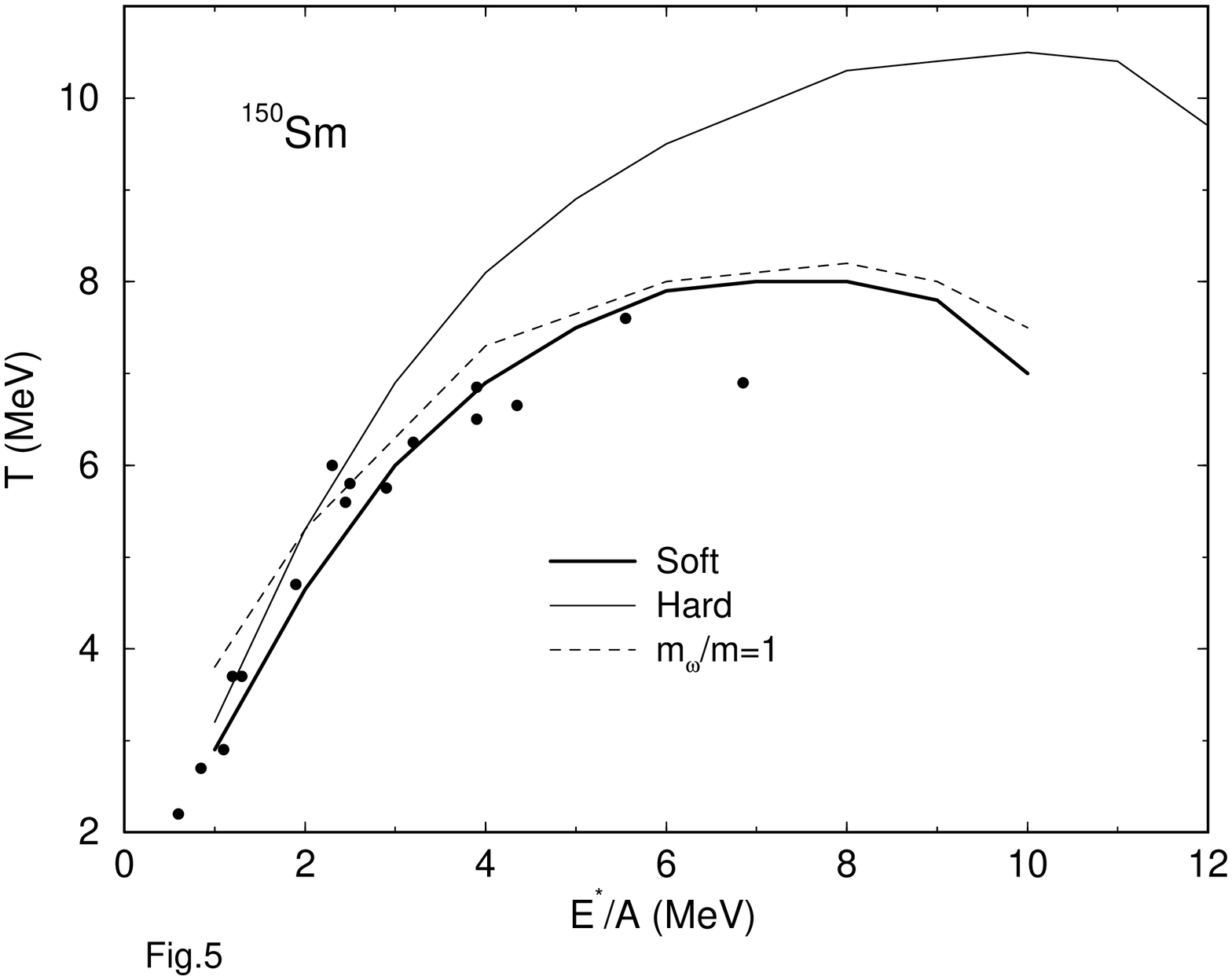}
\end{figure}

\begin{figure}[b]
\includegraphics[width=0.65\textwidth,angle=270,clip=false]{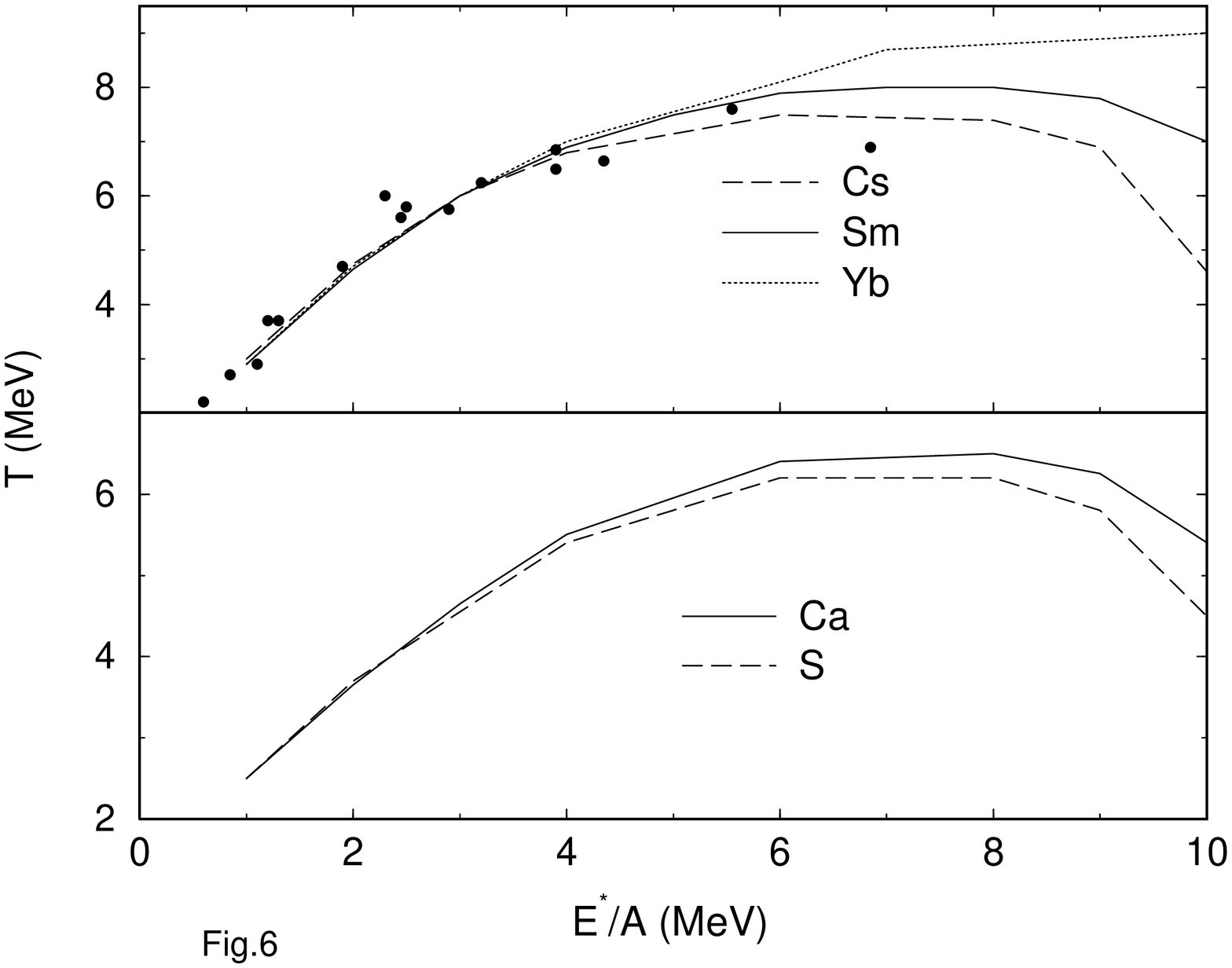}
\end{figure}

\vspace*{-1cm}
\begin{figure}[t]
\includegraphics[width=0.65\textwidth,angle=270,clip=false]{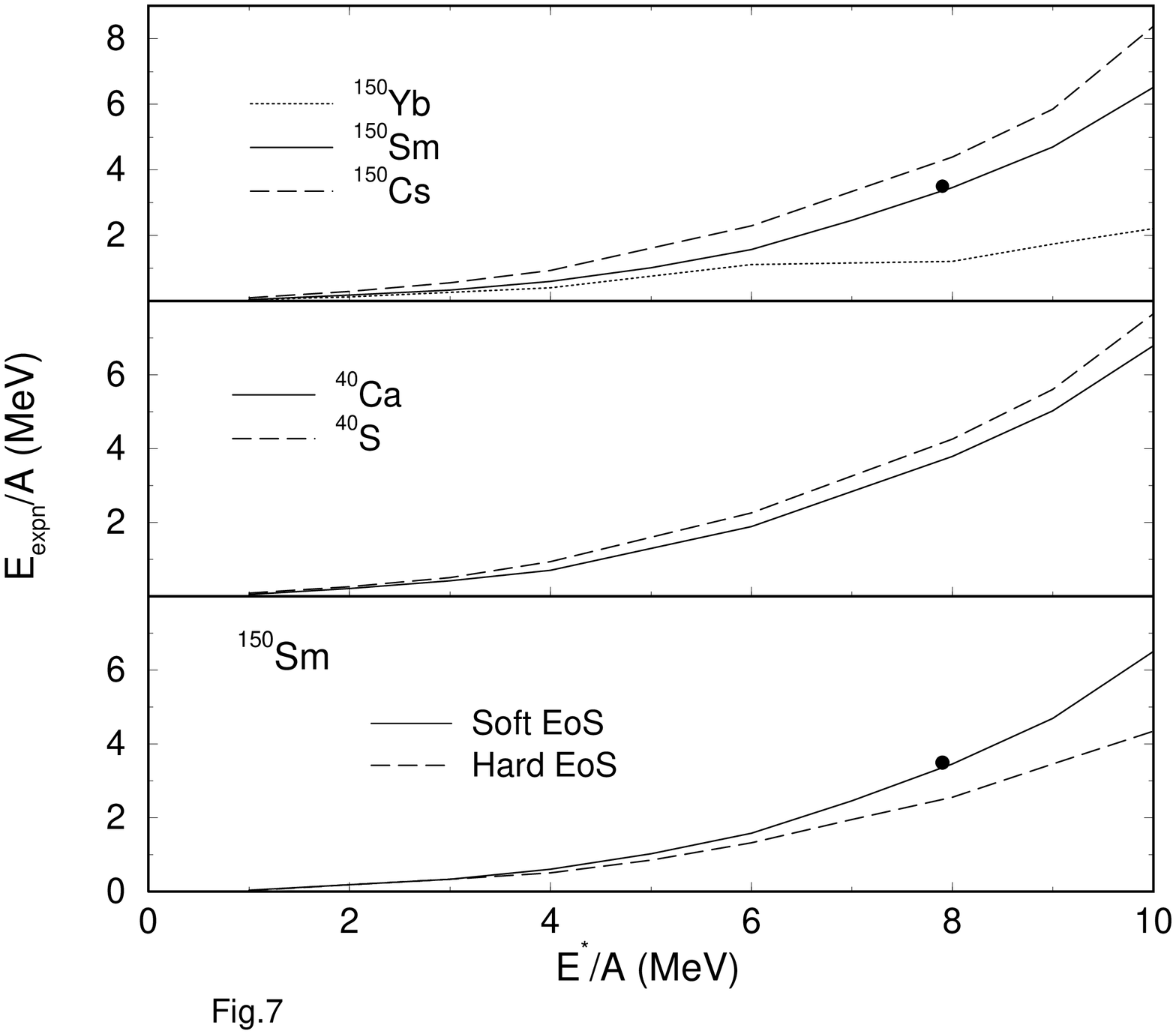}
\end{figure}

\begin{figure}[b]
\includegraphics[width=0.65\textwidth,angle=270,clip=false]{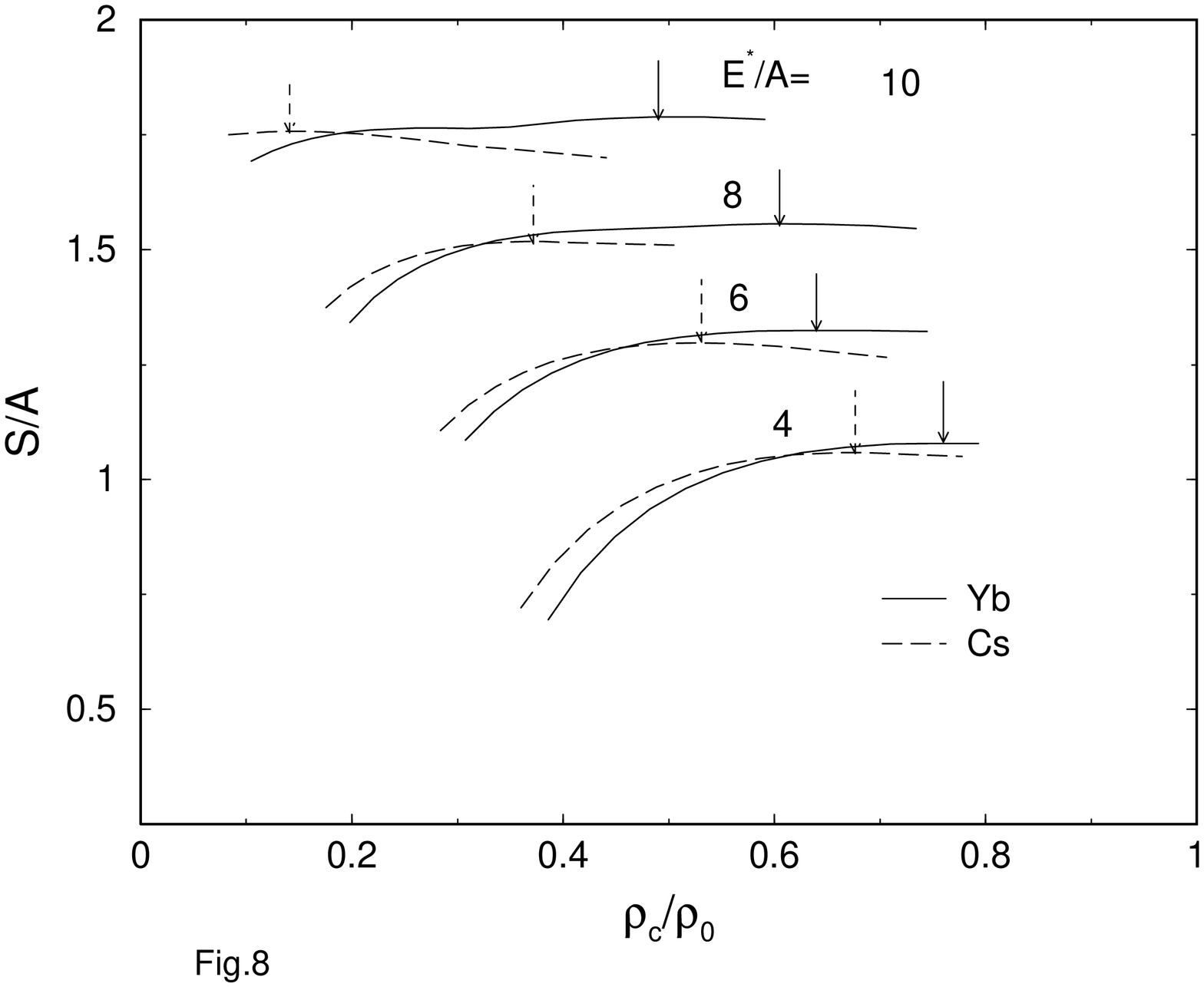}
\end{figure}

\vspace*{-1cm}
\begin{figure}[t]
\includegraphics[width=0.65\textwidth,angle=270,clip=false]{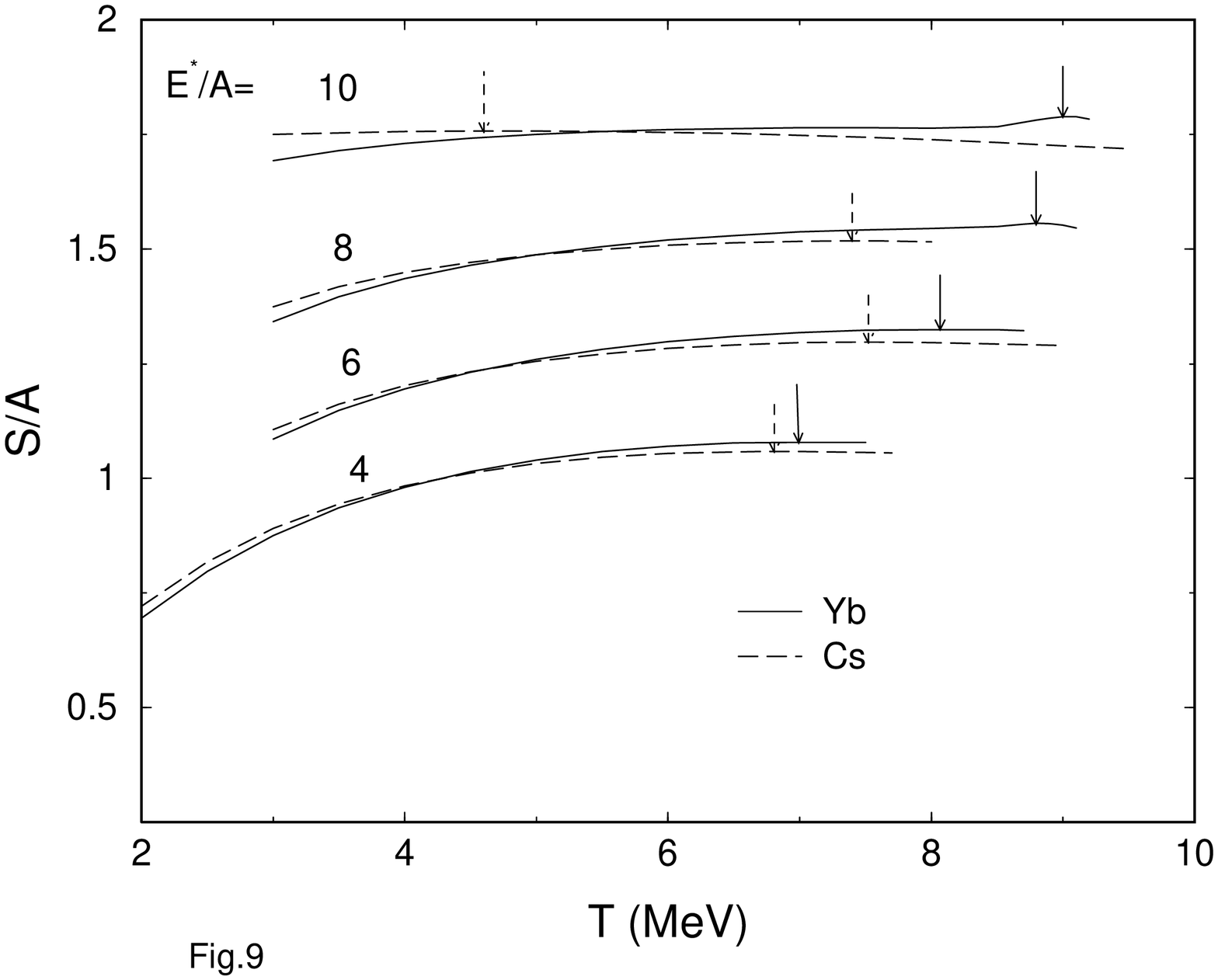}
\end{figure}

\begin{figure}[b]
\includegraphics[width=0.65\textwidth,angle=270,clip=false]{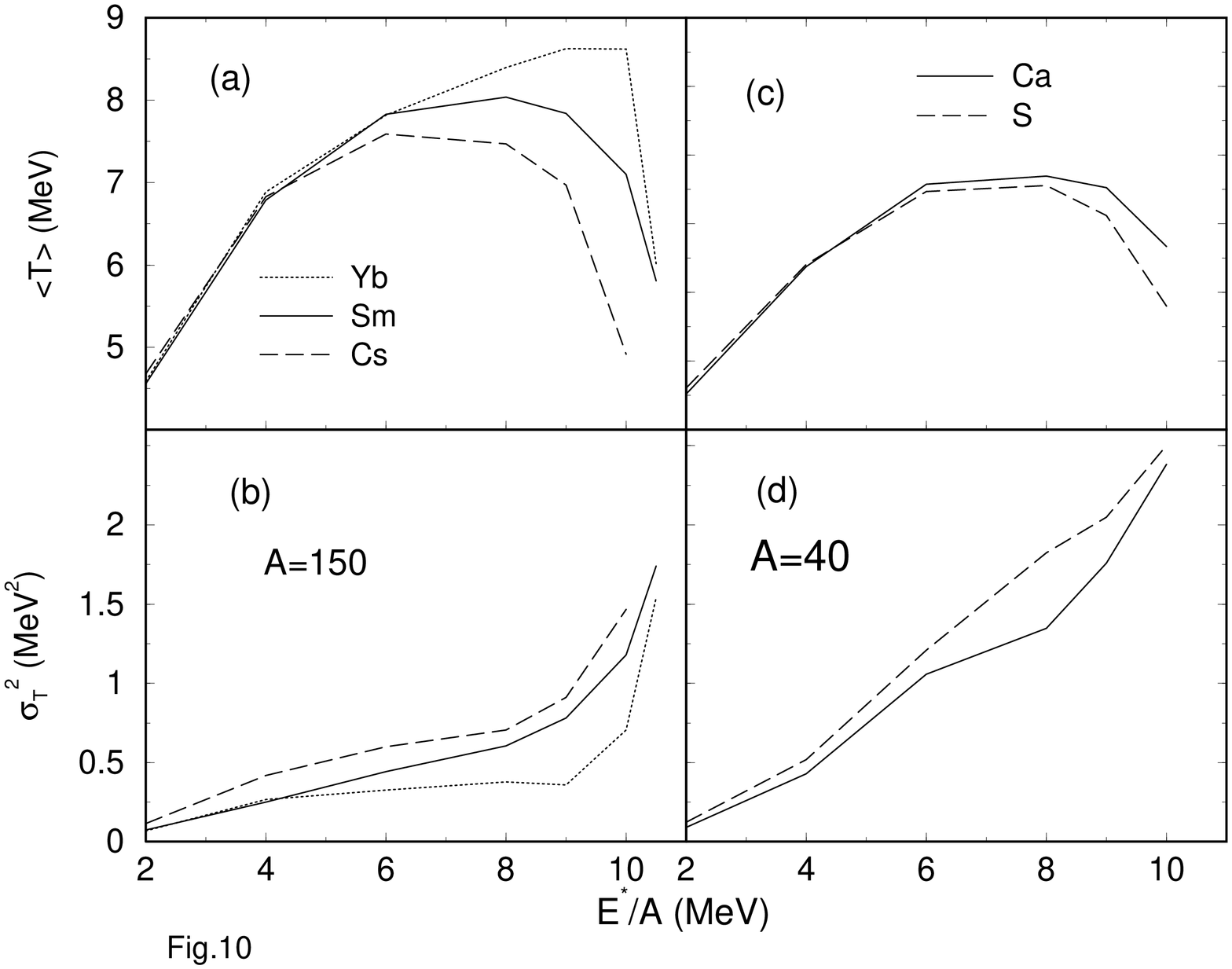}
\end{figure}

\vspace*{-1cm}
\begin{figure}[t]
\includegraphics[width=0.65\textwidth,angle=270,clip=false]{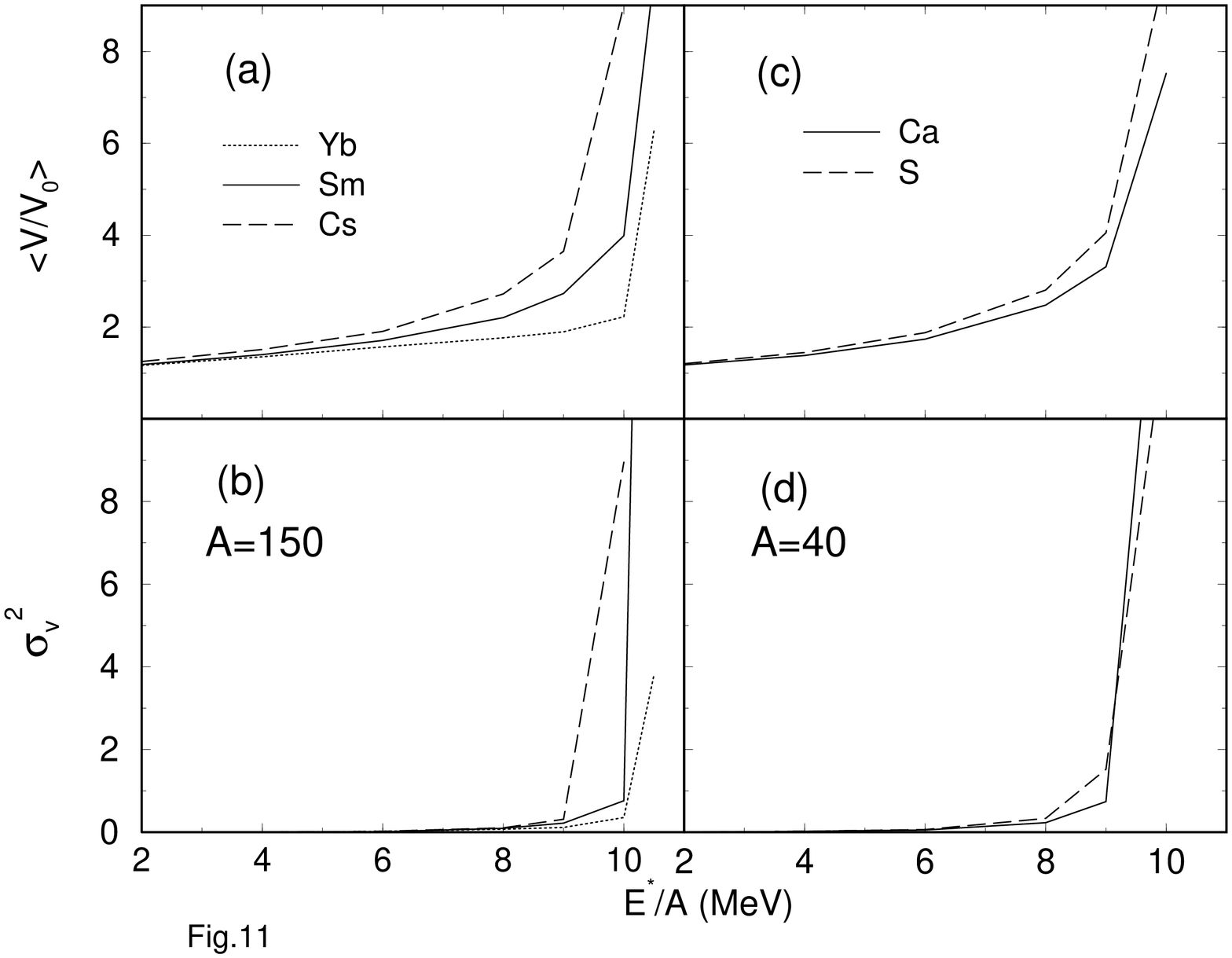}
\end{figure}

\begin{figure}[b]
\includegraphics[width=0.65\textwidth,angle=270,clip=false]{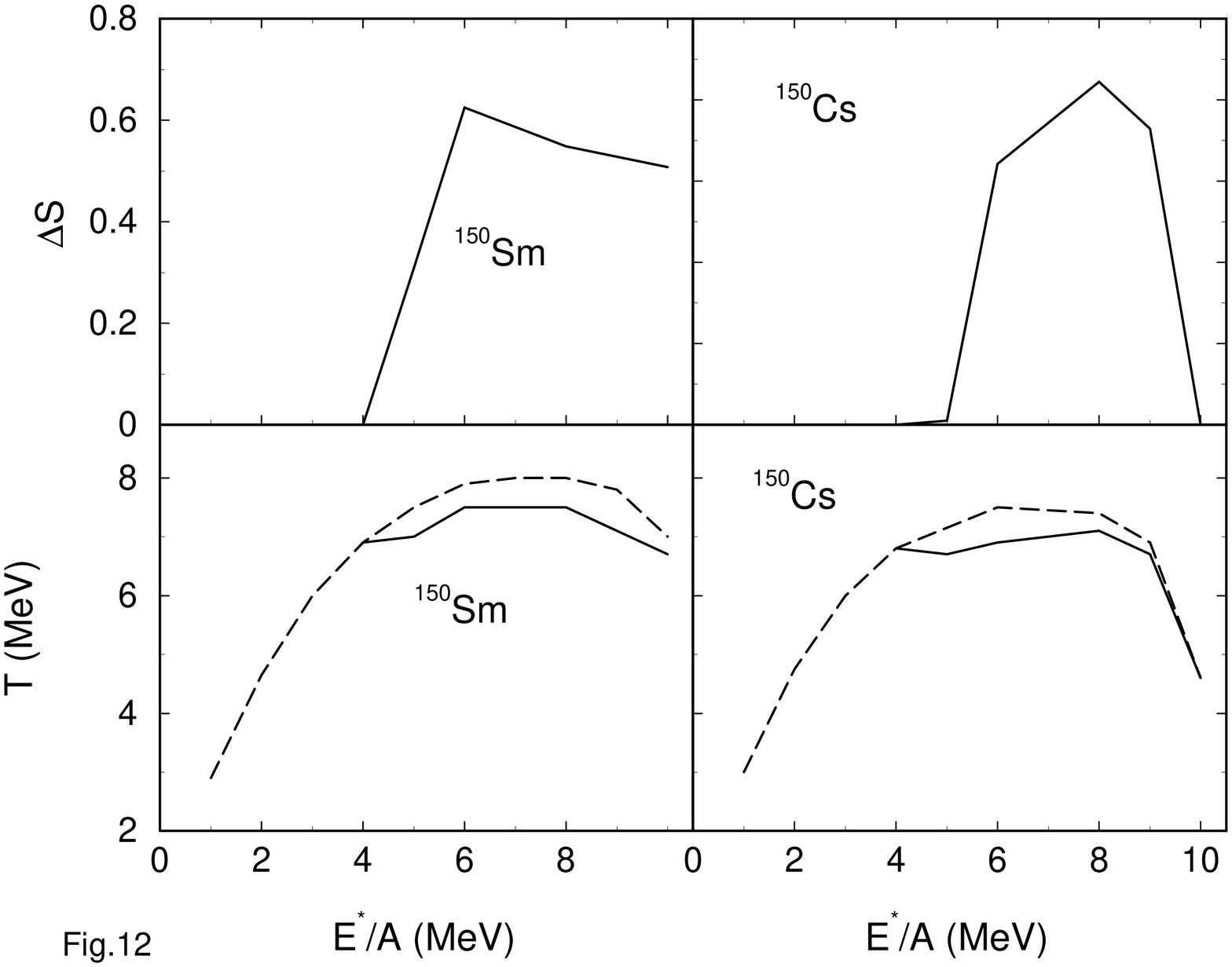}
\end{figure}


\begin{thebibliography}{99}
   
\bibitem{ish} C. Ishizuka, A. Ohnishi, and K. Sumiyoshi,
Nucl. Phys. {\bf A723}, 517 (2003).

\bibitem{bot} A. S. Botvina, and I. N. Mishustin, Phys. Lett.
{\bf B584}, 233 (2004).

\bibitem{li} B. A. Li, and L. W. Chen, Phys. Rev. C {\bf 72}, 064611 (2005).

\bibitem{she} D. V. Shetty, S. J. Yennello, and G. A. Souliotis,
nucl-ex/0610019 (2006).

\bibitem{fri} S. Fritz {\it et al}, Phys. Lett. {\bf B461}, 315 (1999).

\bibitem{vio} V. E. Viola,
 K. Kwiatkowski, J. B. Natowitz, and S. J. Yennello,
Phys. Rev. Lett. {\bf 93}, 132701 (2004).

\bibitem{rad} Ad. R. Raduta {\it et al}, Phys. Lett. 
{\bf B623}, 43 (2005).

\bibitem{nat} J. B. Natowitz {\it et al}, Phys. Rev.C 
{\bf 66}, 031601(R) (2002).

\bibitem{nor} W. Norenberg, G. Papp, and P. Rozmej, GSI preprint
2002-03, January, 2002B.

\bibitem{bon} J. P. Bondorf {\it et al},
Phys. Rep. {\bf 257}, 133 (1995).

\bibitem{gro} D. H. E. Gross, Rep. Prog. Phys. {\bf 53}, 1122 (1990).

\bibitem{das} S. Das Gupta, and J. Pan, Phys. Rev.C 
{\bf 53}, 1319 (1996).

\bibitem{cho} Ph. Chomaz, and F. Gulminelli, Phys. Lett. 
{\bf B447}, 221 (1999).

\bibitem{sob} L. G. Sobotka, R. J. Charity, J. T\~oke, and W. U.
Schr\"oder, Phys. Rev. Lett. {\bf 93}, 132702 (2004).

\bibitem{de} J. N. De, S. K. Samaddar, X. Vi\~nas, and M. Centelles,
Phys. Lett. {\bf B638}, 160 (2006).

\bibitem{sob1} L. G. Sobotka, and R. J. Charity, Phys. Rev. C {\bf 73},
 014609 (2006).

\bibitem{lom} U. Lombardo, and G. Russo, Phys. Rev. C {\bf 36}, 841 (1987).

\bibitem{dav} K. T. R. Davies, and S. E. Koonin, Phys. Rev. C
{\bf 23}, 2042 (1981).

\bibitem{ban} D.Bandyopadhyay, C. Samanta, S. K. Samaddar, and J. N. De,
Nucl. Phys. {\bf A511}, 1 (1990).

\bibitem{de1} J. N. De, N. Rudra, S. Pal, and S. K. Samaddar, Phys. Rev. C
{\bf 53}, 780 (1996). 

\bibitem{uma} V. S. Uma Maheswari, D. N. Basu, J. N. De, and S. K. Samaddar, 
Nucl. Phys. {\bf A615}, 516 (1997).

\bibitem{rud} N. Rudra and J. N. De, Nucl. Phys. {\bf A545}, 608
(1992).

\bibitem{fri1} B. Friedman and V. R. Pandharipande, Nucl. Phys.
{\bf A361}, 502 (1981).

\bibitem{wir} R. B. Wiringa, V. Fiks, and A. Fabrocini, Phys. Rev. C
{\bf 38}, 1010 (1988).

\bibitem{bonc} P. Bonche, S. Levit, and D. Vautherin, Nucl. Phys.
{\bf A436}, 265 (1985).

\bibitem{sur} E. Suraud, Nucl. Phys. {\bf A462}, 109 (1987).

\bibitem{boh} A. Bohr and B. R. Mottelson, {\it Nuclear Structure,
Vol. II} (Benjamin, Reading, MA, 1975).

\bibitem{has} R. Hasse and P. Schuck, Phys. Lett. {\bf B179}, 313 (1986).

\bibitem{bor} P. F. Bortignon and C. H. Dasso, Phys. Lett. {\bf B189},
381 (1987).

\bibitem{pra} M. Prakash, J. Wambach, and Z. Y. Ma,
Phys. Lett. {\bf B128}, 141 (1983).

\bibitem{shl} S. Shlomo and J. B. Natowitz, Phys. Lett. {\bf B252}, 187
(1990).

\bibitem{de2} J. N. De, S. Shlomo, and S. K. Samaddar, Phys. Rev. C
{\bf57}, 1398 (1998).

\bibitem{eis} J. M. Eisenberg and W. Greiner, Microscopic theory
of the nucleus (North Holland, Amsterdam, 1972), p.~394.

\bibitem{mye} W. D. Myers and W. J. Swiatecki, Ann. Phys. {\bf 55},
395 (1969).

\bibitem{rav} D. G. Ravenhall, C. J. Pethick, and J. M. Lattimer,
Nucl.Phys. {\bf A407}, 571 (1983).

\bibitem{bla} J. P. Blaizot, Phys. Rep. {\bf 64}, 171 (1980).

\bibitem{tre} J. Treiner, H. Krivine, O. Bohigas, and J. Martorell,
Nucl. Phys. {\bf A371}, 253 (1981).

\bibitem{maj} M. M. Majumdar, S. K. Samaddar, N. Rudra, and J. N. De,
Phys. Rev. C {\bf 49}, 541 (1994). 

\bibitem{cib} J. Cibor {\it et al.,} Phys. Lett. {\bf B473},
29 (2000).

\bibitem{frie} W. A. Friedman, Phys. Rev. Lett. {\bf 60}, 2125 (1988).

\bibitem{poc} J. Pochodzalla {\it et al.,} Phys. Rev. Lett. {\bf 75},
1040 (1995).

\bibitem{dago} M. D'Agostino {\it et al.,} Phys. Lett. {\bf B473},
219 (2000). 

\end{thebibliography}
\end{document}